\documentclass[usegraphicx,useAMS,usenatbib]{mn2e}
\usepackage{xspace}
\usepackage{amsmath}
\usepackage{verbatim}
\usepackage[normalem]{ulem}
\usepackage{amsfonts,amsmath,amssymb}
\usepackage{times}
\usepackage[italic]{mathastext}
\usepackage{enumitem}
\usepackage[colorlinks,linkcolor=blue,citecolor=blue,urlcolor=blue,breaklinks]{hyperref}
\usepackage{color}
\usepackage{upgreek}
\usepackage{flushend}
\usepackage{lineno}
\usepackage{siunitx}
\usepackage{caption}

\usepackage[dvipsnames]{xcolor}

\def\reff@jnl#1{{\rm#1\/}}
\def\aj{\reff@jnl{AJ}}         
\def\araa{\reff@jnl{ARA\&A}}      
\def\apj{\reff@jnl{ApJ}}        
\def\apjl{\reff@jnl{ApJ}}        
\def\apjs{\reff@jnl{ApJS}}       
\def\aap{\reff@jnl{A\&A}}        
\def\aapr{\reff@jnl{A\&A~Rev.}}     
\def\aaps{\reff@jnl{A\&AS}}       
\def\mnras{\reff@jnl{MNRAS}}      
\def\physrep{\reff@jnl{Physics Reports}}
\def\prd{\reff@jnl{Phys.Rev.D}}     
\def\prl{\reff@jnl{Phys.Rev.Lett}}   
\def\pasp{\reff@jnl{PASP}}       
\def\pasj{\reff@jnl{PASJ}}       
\def\nat{\reff@jnl{Nature}}       
\def\jcap{\reff@jnl{JCAP}}   
\def\memsai{\reff@jnl{MemSAI}} 
\def\na{\reff@jnl{New Astronomy}}       
\def\procspie{\reff@jnl{SPIE}}       
\def\pasa{\reff@jnl{PASA}}       

\def\Cref#1{Chapter~\ref{#1}\xspace}

\usepackage{eso-pic}

\AddToShipoutPictureBG*{
  \AtPageUpperLeft{
    \hspace{10pt}
    \raisebox{-5\baselineskip}{
      \makebox[0pt][l]{\textnormal{DES-2021-0635}}
}}}

\AddToShipoutPictureBG*{
  \AtPageUpperLeft{
    \hspace{10pt}
    \raisebox{-6\baselineskip}{
      \makebox[0pt][l]{\textnormal{FERMILAB-PUB-21-226-AE}}
}}}

\title[The mass and galaxy distribution around SZ-selected clusters]{The mass and galaxy distribution around SZ-selected clusters}

\author[Shin et al.]
{\parbox{\textwidth}{
\Large
T.~Shin,$^{1}$\thanks{E-Mail: taeshin@sas.upenn.edu}
B.~Jain,$^{1}$
S.~Adhikari,$^{2,3}$
E.~J.~Baxter,$^{4}$
C.~Chang,$^{2,3}$
S.~Pandey,$^{1}$
A.~Salcedo,$^{5}$
D.~H.~Weinberg,$^{5}$ 
A.~Amsellem,$^{2}$ 
N.~Battaglia,$^{6}$
% ...
  \vspace{0.2cm}\\
  \parbox{\textwidth}{
(affiliations are listed at the end of the paper)\\}}
}

\begin{document}
\date{\today}
\pagerange{\pageref{firstpage}--\pageref{lastpage}}
\pubyear{2020}
\maketitle
\label{firstpage}
\begin{abstract}
We present measurements of the radial profiles of the mass and galaxy number density around Sunyaev-Zel'dovich (SZ)-selected clusters using both weak lensing and galaxy counts. The clusters are selected from the Atacama Cosmology Telescope (ACT) Data Release 5 (DR5) and the galaxies from the Dark Energy Survey (DES) Year 3 dataset. With signal-to-noise of 62 (43) for galaxy (weak lensing) profiles over scales of about $0.2-20h^{-1}$ Mpc, these are the highest precision measurements for SZ-selected clusters to date.  Because SZ selection closely approximates mass selection, these measurements enable several tests of theoretical models of the mass and light distribution around clusters. Our main findings are: 1. The splashback feature is detected at a consistent location in both the mass and galaxy profiles and its location is consistent with predictions of cold dark matter N-body simulations. 2. The full mass profile is also consistent with the simulations; hence it can constrain alternative dark matter models that modify  the mass distribution of clusters. 3. The shapes of the galaxy and lensing profiles are remarkably similar for our sample over the entire range of scales, from well inside the cluster halo to the quasilinear regime. This can be used to constrain processes such as quenching and tidal disruption that alter the galaxy distribution inside the halo, and scale-dependent features in the transition regime outside the halo. We measure the dependence of the profile shapes on the galaxy sample, redshift and cluster mass.  We extend  the  Diemer \& Kravtsov model for the cluster profiles to the linear regime using perturbation theory and show that it provides a good match to the measured profiles. We also compare the measured profiles to predictions of the standard halo model and simulations that include hydrodynamics. Applications of these results to cluster mass estimation and cosmology are discussed. 
\end{abstract}


\section{Introduction}

Galaxy clusters are the largest gravitationally bound objects in the universe. Their virialization is not considered complete as most clusters are actively accreting matter even at the present epoch.
Clusters have a rich merging history and an anisotropic structure which makes the definition of their halo boundary challenging. Nevertheless the averaged profiles of a large sample of clusters are smooth and isotropic. 
The splashback radius refers to a sharp drop in the mass density profile of dark matter haloes, near the first apocenters of infalling matter.
It was first identified in $N$-body simulations by  \citet{Diemer2014} via stacked mass profiles of haloes at different redshifts and stages of evolution and investigated in several further studies \citep[e.g.,][]{Adhikari14,More2015,Xun16,Mansfield2017,Diemer2017}. Its application to data holds great promise for  astrophysical and cosmological studies with clusters. 

The density profiles of cluster haloes can be probed observationally in several ways, for example, by studying
the distribution of galaxies in haloes, or by stacked measurements of the weak gravitational lensing of background galaxies around haloes to get the matter distribution essentially directly. \citet{More16} used the projected galaxy number density profile around redMaPPer \citep[RM;][]{Rykoff2014} galaxy clusters identified in data from the Sloan Digital Sky Survey \citep[SDSS;][]{Aihara2011} to present the first evidence for a splashback feature. Subsequently, evidence for the feature was established by \citet{Baxter17} using the projected galaxy number density profiles around two samples of SDSS-identified clusters, and by \citet{Chang2017} using the galaxy density and weak lensing profiles around RM clusters identified in the first year of Dark Energy Survey (DES) data (see also  \citealt{Umetsu17}, \citealt{contigiani19}, \citealt{Murata20} and \citealt{Bianconi20}). 
Note that \citet{Tomooka20} also measured a sharp radial transition at the edge of galaxy clusters in the line of sight velocity dispersion of tracer galaxies around RM clusters in the SDSS spectroscopic survey.
Also note that splashback-like features in individual clusters were discussed in previous studies \citep{Rines2013,Tully2015,Patej2015}.

In all of the cases involving photometric surveys, the evidence for the splashback feature came from identifying the presence of a sharp steepening in the projected halo (galaxy/dark matter) density profiles. Interestingly, for clusters identified via the RM algorithm, and for measurements using the projected galaxy number density profile around clusters, the location of splashback is about 20\% ($\sim 3\sigma$) smaller than predictions from $N$-body simulations \citep{More16,Baxter17,Chang2017}.

\citet{Busch2017}, \citet{Zu2017} and subsequent work explored whether cluster-finding algorithms like RM can imprint artificial splashback-like features into cluster density profiles via selection effects.
They pointed out that the chance projection of galaxies, especially in the cluster outskirts, can affect the measured galaxy density profile, biasing the measurements made with optically selected clusters. 
\citet{Zu2017} also showed that  projection preferentially occurs in dense regions, causing a correlation between the large scale overdensity and the concentration inferred with the member galaxies. 
\citet{Chang2017} investigated possible systematics involved with the RM cluster-finding algorithm by varying the member-searching aperture around clusters. They found that the location of the splashback was somewhat sensitive to the size of the aperture. \citet{Murata20} examined the splashback feature from an independent optical cluster finder in the Hyper Supreme-Cam \citep{Aihara2017} data and found the splashback feature to be more consistent with simulations than RM.

To avoid these complexities of optically selected clusters, in \citet{Shin19} we  used SZ-selected clusters correlated with galaxies and found no evidence for selection artifacts. We found that the splashback radius measured around SZ-selected clusters is consistent with $N$-body simulations of cold dark matter (CDM) \citep[see also][]{Zuercher19}.
Finally, in  previous studies with SZ-selected clusters \citep{Shin19, Adhikari2020}, we also studied the density profiles of galaxies split by galaxy colors, extending the formalism in \citet{Baxter17,Chang2017} and used it to constrain the quenching timescales of the galaxy star formation inside the clusters.

In this paper we use SZ-selected clusters and measure the mass density profile and the splashback radius around SZ-selected galaxy clusters using weak gravitational lensing. 
The goal of this work is to measure the projected radial mass density profile using weak lensing and compare it with the projected galaxy number density distribution  and theoretical predictions for these profiles. Our analysis relies on a catalog of galaxy clusters \citep{Hilton_2020} that have been observed via their SZ signal in millimeter-wave maps from the ACT Data Release 5 \citep[ACT DR5;][]{Naess_2020} by the Atacama Cosmology Telescope \citep[ACT,][]{Fowler07, Thornton16}. 
SZ selection is essentially redshift independent, and appears closer to halo mass selection than  optical cluster finders because of the smaller scatter in the relationship between cluster mass and observable and lower bias over the mass range probed so far. 
Moreover, clusters selected with the SZ effect suffer less from systematic effects such as line-of-sight projections and triaxiality than optically selected clusters \citep[e.g.,][]{Shin19}. As a caveat, we note that detailed studies with mock cluster catalogs from simulated SZ maps are still needed to confirm these conclusions. We cross-correlate the cluster positions with galaxies from DES Year 3 (DESY3; the data taken in the first 3 years of the survey) data and with the lensing shear measured from background galaxies in the same dataset \citep{DES2005,y3gold,Y3shape}.

Weak lensing gives us a direct probe of the matter distribution and the gravitational potential of the cluster halo that is traced by the visible galaxies. Therefore, the comparison between the galaxy number density profile and mass distribution opens up the opportunity to study how processes that exclusively affect galaxies, as, for example, tidal disruptions, harassment and ram-pressure stripping alter their relative clustering. Comparison between the matter and galaxy distributions can also help understand the nature of gravity \citep{Schmidt10, Adhikari2018, Contigiani:2018hbn}.  Direct comparison of the  mass density profile measured through weak lensing  with predictions from $N$-body simulations of CDM and hydrodynamical simulations will allow us to understand the impact of baryonic physics on cluster profiles. We can constrain the effects of dark matter interactions by studying the small central region where cores are expected to form in certain dark matter models \citep[see][for a review]{Buckley18}  and also beyond it, in the outskirts where the matter profile can be significantly steeper than in CDM \citep{Banerjee20}. Finally, \citet{Xhakaj20} also show that measuring the mass distribution directly from lensing can help constrain cluster accretion rates. 

Here we present the first simultaneous measurements of the mass density and the galaxy number density profiles of SZ-selected clusters from the ACT DR5 \citep{Hilton_2020} and make some initial comparisons with CDM and hydrodynamical simulations.
The paper is organized as follows. The input data catalogs from DES and ACT are described in Section 2. Modeling of measurements of the profiles is presented in Section 3 and the interpretation of the results in Section 4. We conclude in Section 5. 
Throughout this paper, $r$ represents 3D halo-centric radius, while $R$ represents the projected 2D radius.

\section{Data}
\label{sec:data}

\subsection{The ACT DR5 SZ-selected cluster sample}
\label{sec:data-szcl}

The cluster sample used in this study is selected from the ACT DR5 cluster catalog \citep{Hilton_2020}, which consists of 4195 SZ-selected galaxy clusters, detected with S/N~$> 4$, in a survey area of 13,211\,deg$^2$. 
The clusters were detected by applying a multi-frequency matched filter to 98 and 150\,GHz maps, constructed from ACT observations obtained from 2008-2018 (see \citealt{Naess_2020} for details of the map making procedures). 
Optical confirmation, removal of false detections and redshifts for the ACT DR5 clusters come from a variety of large area optical/IR surveys from which the locations of the brightest cluster galaxies (BCGs) are determined, with 4600\,deg$^2$ covered by DES.
We use the locations of the BCGs as the cluster centers for our calculation.
\footnote{Note that approximately 35\% of the BCG locations are determined by visual inspection, while the remaining are from the DES \textsc{redMaPPer} clusters \citep{McClintock19}.}

\begin{figure}
\centering
\includegraphics[width=0.99\linewidth]{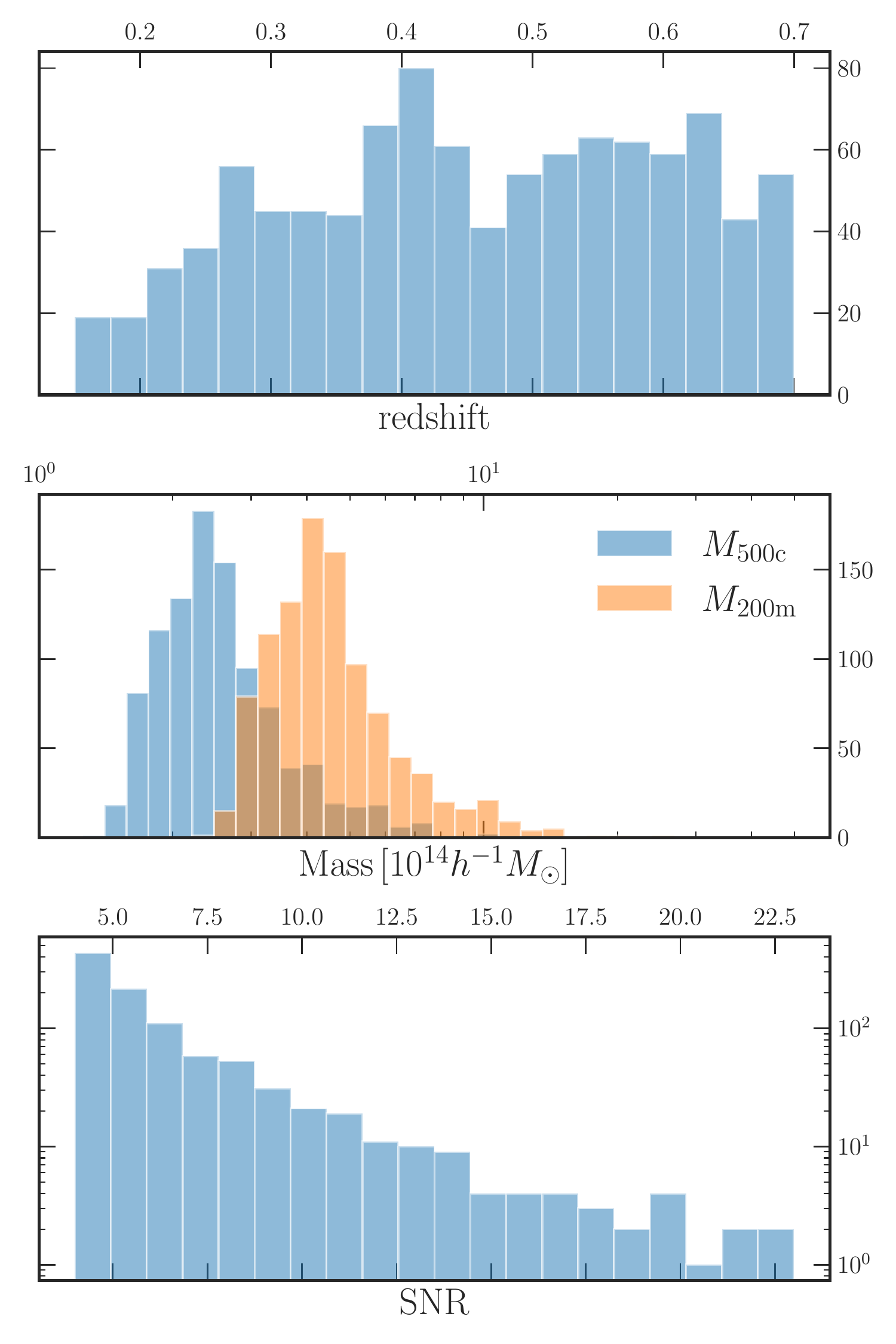}
\caption{The distributions of the cluster sample over redshift (top), estimated cluster mass (middle) and S/N (bottom). Two definitions of cluster mass are shown in the middle panel with the richness-based WL mass correction applied, see Sec.~\ref{sec:data-szcl} and \ref{sec:model-fitting} for details. 
}
\label{fig:cl-dist}
\end{figure}

In this work we use the clusters with redshifts between 0.15 and 0.7, and S/N~$> 4$, and select those that lie inside the DESY3 footprint. 
There are 1002 clusters in the resultant fiducial cluster sample.
We show the redshift,  mass and  S/N distributions of our cluster sample in  Fig.~\ref{fig:cl-dist}.
The mean mass and the redshift of the sample are $M_{\rm 500c} = 2.72 \times 10^{14} h^{-1} M_{\odot}$ and 0.46, respectively, where $M_{\rm 500c}$ is the mass inside the radius at which the average interior  density is 500 times the critical density, $\rho_{\rm c}$.
We make use of the mass estimates inferred from the cluster SZ signals that have
been rescaled according to a richness-based weak-lensing mass calibration procedure, as described in \citet{Hilton_2020}. 
To facilitate comparison with other work, we also show the distribution of $M_{\rm 200m}$ in the figure, where $M_{\rm 200m}$ is the mass inside the radius at which average interior density equals 200 times the mean mass density of the universe. Our stacked cluster profiles average over all masses in the sample. Since the mass function steeply falls with halo mass for such massive clusters, we have checked with simulations that features in the profile are barely degraded by the averaging and recover the profile of the mean mass of the sample. As a test with the data, we also show results from splitting the sample on estimated cluster mass and redshift (see Appendix A). 

For some applications we make use of results from mock ACT DR5 cluster catalogs as random points.
These are generated by sampling from the \citet{Tinker2008} mass function, applying an SZ-signal--mass relation, and comparing the predicted signals with randomly chosen positions in the SZ-signal noise map generated by the cluster finder. 
The scaling relation parameters are chosen such that the number of S/N~$> 6$ clusters in these mock catalogs approximately matches the number of clusters observed in the real data within
the DES footprint, where the real sample is approximately 100\% pure and complete in terms of
redshift follow-up \citep[see Section~3.3 of][]{Hilton_2020}.

\subsection{DES Year 3 galaxy catalog}
\label{sec:data-desgold}

The galaxies and their shapes to be correlated with our cluster sample are obtained from the DESY3 data. 
DES \citep{DES2005} is a multi-color imaging survey covering $\sim$5000 square degrees of the South Galactic Cap.
Using the 570-megapixel Dark Energy Camera \citep{Flaugher2015} mounted on the Cerro Tololo Inter-American Observatory (CTIO) 4m Blanco telescope in Chile, it images the sky in  \textit{grizY} filters. 
The DES Data Management (DESDM) system \citep{Sevilla2011,Morganson2018} processes the raw images; the high-quality galaxy catalog (Y3GOLD) is generated \citep{y3gold} after careful analysis of the galaxy images including image detrending and processing, photometric calibration, and object classification.
The final products of Y3GOLD comprises 390 million galaxies reaching up to $i_{\rm AB} \sim 23$ at ${\rm S/N} \sim 10$.
In this analysis, we make use of the Single Object Fitting photometry for which we refer readers to \citet{y3gold}.

When calculating the  galaxy density profile (Sec.~\ref{sec:measure-dsigma}), following \citet{Shin19} and \citet{Adhikari2020}, we apply further selection criteria to the galaxies after filtering out stars and photometric failures:  apparent magnitude cut $m_i < 22.5$, and color cuts  $-1 < m_g - m_r < 3,\, -1 < m_r - m_i < 2.5\,$ and $-1 < m_i - m_z < 2$  to exclude galaxies with extreme colors that may result in catastrophic failures in photo-$z$ estimation \citep{crocce2018}.
We further require  galaxies to have errors in the \textit{i}-band magnitude smaller than 0.1. 
To ensure the completeness of the galaxy sample we select a sky footprint that is sufficiently deep for our limiting magnitude cut $m_i=22.5$, 
which gives a sky area of $\sim$4460 square degrees.
The resulting galaxy catalog that we use in this study contains about 84 million galaxies.

When calculating the galaxy profile at different redshifts (Sec.~\ref{sec:measure-dsigma}), we require that the \textit{i}-band absolute magnitude ($M_i$) is smaller than -19.87 (which corresponds to the apparent magnitude cut ($m_i<22.5$) at the maximum redshift used, $z=0.7$) using distance modulus.
The absolute magnitude cut ensures greater consistency of the galaxy sample at different redshifts (see Sec.~\ref{sec:systematics} for a caveat on this).

\subsection{DES Year 3 galaxy shape catalog}
\label{sec:data-shear}

The weak lensing shape catalog associated with the Y3GOLD galaxies is obtained using the \textsc{Metacalibration} algorithm \citep{Huff2017,Sheldon2017} and presented in \citet{Y3shape}.
Here, we provide a brief summary of how  \textsc{Metacalibration} determines the shapes of galaxies and their relationship to the WL shears.

\textsc{Metacalibration} in DESY3 measures the shapes of galaxies in \textit{riz} bands. 
The  DESY3, shear catalog covers 4143 square degrees of the sky, with a source number density of $5.59\, \mathrm{gal/arcmin}^2$ and shape noise of 0.261. 
The shape of a galaxy is defined as a 2-component ellipticity, $\textbf{e} = |e|\exp{2i\phi}$, where $\phi$ is the angle from the x-axis of the coordinate system to the major axis of the galaxy.
We refer readers to \citet{Y3shape} for details.
After the cuts on signal-to-noise ratio, size, and the magnitude, the final number of the galaxies that have passed these cuts is about 100 million.

\textsc{Metacalibration} relates the measured shapes of the galaxies, \textit{\textbf{e}}, to the true shear from gravitational lensing, $\boldsymbol\gamma$, by the response matrix, $\mathcal{R}$. 
That is:
\begin{equation}
    \langle \boldsymbol\gamma \rangle = \langle \mathcal{R} \rangle^{-1} \langle \textbf{e} \rangle,
\end{equation}
where the angled brackets denote the ensemble average. 
In other words, the response $\mathcal{R}$ represents the response of the measured galaxy shapes to the true shear. 
In addition, the galaxy shapes depend on the specific galaxy sample selection.
Therefore, \textsc{Metacalibration} calculates an additional response term, the selection response, $\mathcal{R}_{\rm s}$, which represents the response of measured shapes specific to the selected galaxy sample.
The total response then becomes $\mathcal{R}+\mathcal{R}_{\rm s}$.
Note that the response is a 2$\times$2 matrix but it is well represented by the average of the diagonal components, which we adopt here \cite[following, e.g.,][]{Prat2018}.

Note that we are not considering a few percent level multiplicative shear biases inferred from image simulations in \citet{Maccrann20}. 
These biases mainly arise from blending of source galaxies, which is not fully accounted for by the \textsc{Metacalibration} method. 
However, since the main focus of this study is the shapes of cluster density profiles, such multiplicative biases can be safely overlooked.

\subsection{Photometric redshift}
\label{sec:photoz}

As described in \citet{y3gold}, the redshifts of the galaxies are estimated using the Bayesian Photometric Redshifts (BPZ) algorithm \citep{bpz, Hoyle18} which fits the galaxy magnitudes in \textit{griz} bands to SED templates.
Note that we do not directly use the galaxy photo-$z$'s for our  cross-correlation analyses, therefore the details of BPZ are not of primary interest in this paper.
In Sec.~\ref{sec:theory_comp}, we do use the galaxy redshift distribution to estimate the large scale galaxy bias that normalizes the theoretical predictions. The comparison of theory to measurements has 10 percent level uncertainties due to a combination of factors that include photo-$z$'s.
The uncertainty in the photo-$z$ estimates could also induce a multiplicative bias in the measured WL profile \citep{McClintock19}, which could bias the mass estimation at a level well below the statistical error (in particular in the comparison of the inferred splashback radius to theory).
The main focus of this study is on the shapes of the cluster density profiles which are immune to such multiplicative biases.

\subsection{Simulations}
\label{sec:data-sim}

To compare our observational results with DM-only simulations we use particle data from the Multidark Planck (MDPL2) simulation\footnote{\url{https://www.cosmosim.org/cms/simulations/mdpl2/}} \citep{Klypin.etal.2016} at $z=0.49$; this is the publicly available snapshot that is closest to the mean redshift (0.46) of our cluster sample. The MDPL2 is an $N$-body simulation with $3840^3$ particles that simulate a $1^3\,h^{-3}\mathrm{Gpc}^3$ volume using the Planck Cosmology \citep{Planck18}. The halo catalogs have been generated using the Rockstar halo finder and the halo histories are generated using the Consistent trees algorithm \citep{Behroozi13}. We select a sample of clusters that match the distribution of halo mass of our sample. We extract particles out to a radius of $50$ Mpc $h^{-1}$ around the halos from a downsampled set of particles at the mean redshift of our cluster sample and compare the measured density profiles to the data. To emulate the accurate redshift weighting of our observed sample, apart from using  MDPL2, we also use a lower resolution simulation with the same volume but with $1024^3$ particles. We use dark matter particle data from 30 snapshots in the interval $0.15 < z < 0.7$ to match the mass and redshift distribution of our simulation clusters to the observed cluster sample.

We also make preliminary comparisons with hydrodynamical simulations. We use the IllustrisTNG simulations \citep{Nelson:2015dga} to study the matter and projected galaxy number density profiles around clusters. IllustrisTNG is a state-of-the-art cosmological, magneto-hydrodynamical simulation that uses the AREPO code \citep{arepo} to evolve a universe with dark matter and baryons. In particular we use the TNG300 simulation that simulates a $300^3\,h^{-3}\mathrm{Mpc}^3$ cosmological volume. Given the smaller volume of the simulation, we do not have a large number of cluster mass halos at the mass range explored in data. Therefore we study all clusters with $M_{200m}>10^{14}$ $M_\odot h^{-1}$ in units of $r/r_{\rm 200m}$, where $r_{200 \rm m}$ is the radius that encloses 200 times the background matter density and $M_{200m}$ is the mass enclosed within it. Our sample has 89 clusters at the mean redshift of the sample, $z=0.49$, which is the closest publicly available snapshot to our mean cluster redshift. We study both the cluster--matter and the cluster--galaxy cross-correlation as described in section \ref{sec:theory_comp}.

\section{Modeling and measurement methodology}
\label{sec:measurement}

\subsection{Models for WL and galaxy number density profiles}
\label{sec:model}

We model the weak lensing and projected galaxy number density profiles by integrating the spherically symmetric 3D cluster density profile along the line of sight.
Our model for the 3D density profile is based on the fitting formula proposed by \citet[][hereafter DK14]{Diemer2014}.   The density profile is written as the sum of two components:
\begin{equation}\label{eq:rhoDK}
\rho(r)= \rho_{\rm coll}(r)+\rho_{\rm outer}(r)    
\end{equation}
where,
\begin{equation}
    \rho_{\rm coll}(r) = \rho_{\rm inner}(r)f_{\rm trans}(r),
\end{equation}
\begin{equation}
    \rho_{\rm inner}(r)= \rho_{\rm s} \exp\left(-\frac{2}{\alpha} 
    \left[\left(\frac{r}{r_{\rm s}}\right)^\alpha-1\right]\right),
\end{equation}
\begin{equation}
    f_{\rm trans}(r)=\left[1+\left(\frac{r}{r_{\rm t}}\right)^\beta \right]^{-\gamma/\beta},
\end{equation}
and 
\begin{equation}
    \rho_{\rm outer}(r)=\rho_0\left[\frac{1}{\tau_{\rm max}}+\left(\frac{r}{r_0}\right)^{s_{\rm e}}\right]^{-1}.
\end{equation}
Here, $\rho_{\rm inner}(r)$ is an Einasto profile \citep{Einasto} truncated by $f_{\rm trans}(r)$ near the splashback radius, representing the contribution from material that is in orbit around the cluster.  
The contribution from nearby matter not in orbit around the cluster is represented by $\rho_{\rm outer}(r)$, which is dominated by infalling matter. Its profile is close to a pure power law, as expected from the spherical collapse model.  Since $s_e > 0$, the quantity $\tau_{\rm max}$ limits the maximum value that $\rho_{\rm outer}$ can reach at the center of the haloes, preventing $\rho_{\rm outer}$ from dominating over $\rho_{\rm inner}$ at small radii \citep{Colossus}.
We set $\tau_{\rm max}=20$, and confirm that the choice of $\tau_{\rm max}$ does not affect the model fitting significantly as long as $\rho_{\rm coll}$ dominates over $\rho_{\rm outer}$ at small radii.
We fix $r_0= 1.5\,h^{-1} {\rm Mpc}$, since it is degenerate with $\rho_0$ at large radii and $\rho_{\rm outer}$ becomes negligible at small radii compared to $\rho_{\rm coll}$.  The free parameters of the model are $\rho_s$, $\alpha$, $r_s$, $r_t$, $\gamma$, $\beta$, $\rho_0$ and $s_e$.

We integrate the 3D profile along the line of sight to compute the projected density profile, $\Sigma(R)$, at projected distance $R$. 
This integral is performed between  $l_{\rm max}=\pm 40 \, h^{-1} {\rm Mpc}$, where $\l$ is the distance along the line of sight to the cluster, with the origin centered on the cluster. It is given by: 
\begin{equation}
    \Sigma_0(R) = \int_{-l_{\rm max}}^{l_{\rm max}} \rho(\sqrt{R^2+l^2}) \, dl.
\end{equation}

So far we have assumed that the true cluster center is known.
However, in practice we take the brightest central galaxy (BCG) as the cluster center, which may not always be the true center of mass.\footnote{Note that we use the position of the BCG as the cluster center, since the 1-2 arcmin resolution of ACT at 150 and 98 GHz  makes the SZ cluster center less reliable.} 
The effect of such miscentering is to modify the density profile, particularly at small radii.  We write the measured density profile as the sum of a miscentered and correctly centered component:
\begin{equation}\label{eq:sigmag_theory}
\Sigma(R) = (1-f_{\rm mis})\Sigma_0(R) + f_{\rm mis}\Sigma_{\rm mis}(R),
\end{equation}
where $\Sigma_{\rm mis}$ is the profile of the miscentered haloes, and $f_{\rm mis}$ the fraction of miscentered haloes. 
Following e.g., \citet{Rykoff2016}, $\Sigma_{\rm mis}(R)$ can be modeled as
\begin{equation}\label{eq:sigmag_wmis_theory}
    \Sigma_{\rm mis} (R) = \int {\rm d}R_{\rm mis}P(R_{\rm mis})\Sigma_{\rm mis}(R|R_{\rm mis}),
\end{equation}
where $P(R_{\rm mis})$ is the probability distribution of a halo to be mis-centered by a distance $R_{\rm mis}$ from the true center and 
\begin{equation}
    \Sigma_{\rm mis}(R|R_{\rm mis}) = \\
    \int^{2\pi}_0 \frac{{\rm d}\theta}{2\pi}\Sigma_0\Big(\sqrt{R^2 + R^2_{\rm mis} + 2RR_{\rm mis}{\rm cos}\theta}\Big),
\end{equation}
is the profile of a halo miscentered by a distance $R_{\rm mis}$.  Assuming the distribution of incorrect centers is a 2D Gaussian, $P(R_{\rm mis})$ is described by a Rayleigh distribution:
\begin{equation}
    P(R_{\rm mis}) = \frac{R_{\rm mis}}{\sigma^2_R} {\rm exp}\left[-\frac{R^2_{\rm mis}}{2\sigma^2_R}\right].
\end{equation}
Following \citet{Rykoff2016}, we assume $\sigma_R=c_{\rm mis} R_{\lambda}$ and $R_{\lambda}=\left(\lambda/100 \right)^{0.2}$, where $\lambda$ is the cluster richness.  Since $R_{\lambda}$ changes slowly with $\lambda$, we can for simplicity evaluate $\sigma_R$ at the mean richness of the sample.  We calculate the mean richness by matching the ACT DR5 clusters to those detected by DES, finding that the mean richness of our cluster sample is $\bar{\lambda} = 66$.
Miscentering introduces two new free parameters --- $f_{\rm mis}$ and $c_{\rm mis}$ --- into our model.  

The weak lensing measurements, $\Delta\Sigma$, are sensitive to the shape of the profile inside the radius of 0.2$h^{-1}{\rm Mpc}$, below which we do not have any measurements.  This is because
\begin{equation}
\label{eq:deltaSigma}
    \Delta\Sigma(R)  = \bar{\Sigma}(<R) - \Sigma(R),
\end{equation}
where
\begin{equation}
    \bar{\Sigma}(<R) \equiv \frac{2}{R^2} \int^R_0 \, dR' \, R' \, \Sigma(R').
\end{equation}
The non-locality of the weak lensing measurements is somewhat undesirable, since it introduces correlation between the model profile at small scales and large scales.
To reduce this, we separate the contribution of $[0,0.2]\,h^{-1}{\rm Mpc}$ from the integral:
\begin{equation}
    \Delta\Sigma(R) =  \frac{2}{R^2} \left( \mu + \int^R_{R_{\rm min}} \, dR' \, R' \, \Sigma(R')\right) - \Sigma(R),
\end{equation}
where $\mu\equiv\int^{R_{\rm min}}_0 \, dR' \, R' \, \Sigma(R')$, $R_{\rm min} = 0.2 h^{-1} {\rm Mpc}$, and we set $\mu$ as an additional free parameter in the model when fitting $\Delta \Sigma$.

\begin{figure}
\centering
\includegraphics[width=0.99\linewidth]{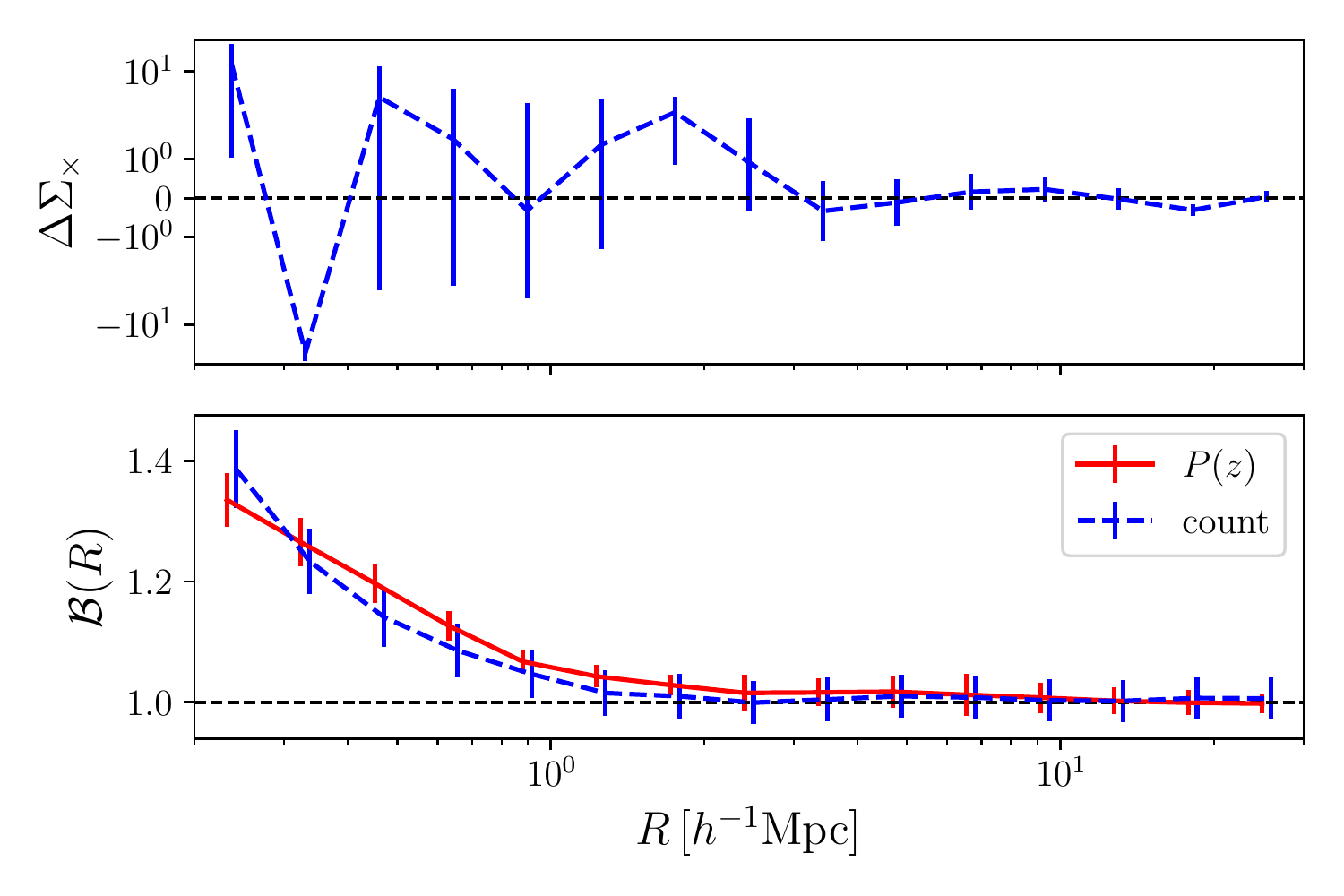}
\caption{
\textit{Upper panel:} The measured cross component from lensing, $\Delta\Sigma_{\rm x}$, which is consistent with zero as expected.
\textit{Lower panel:} The boost factors calculated with two different method: $P(z)$  decomposition (red) and cross-correlation (blue), as described in Section \ref{sec:boost}. 
}
\label{fig:measurement-wl}
\end{figure}

\begin{figure*}
\centering
\includegraphics[width=0.49\linewidth]{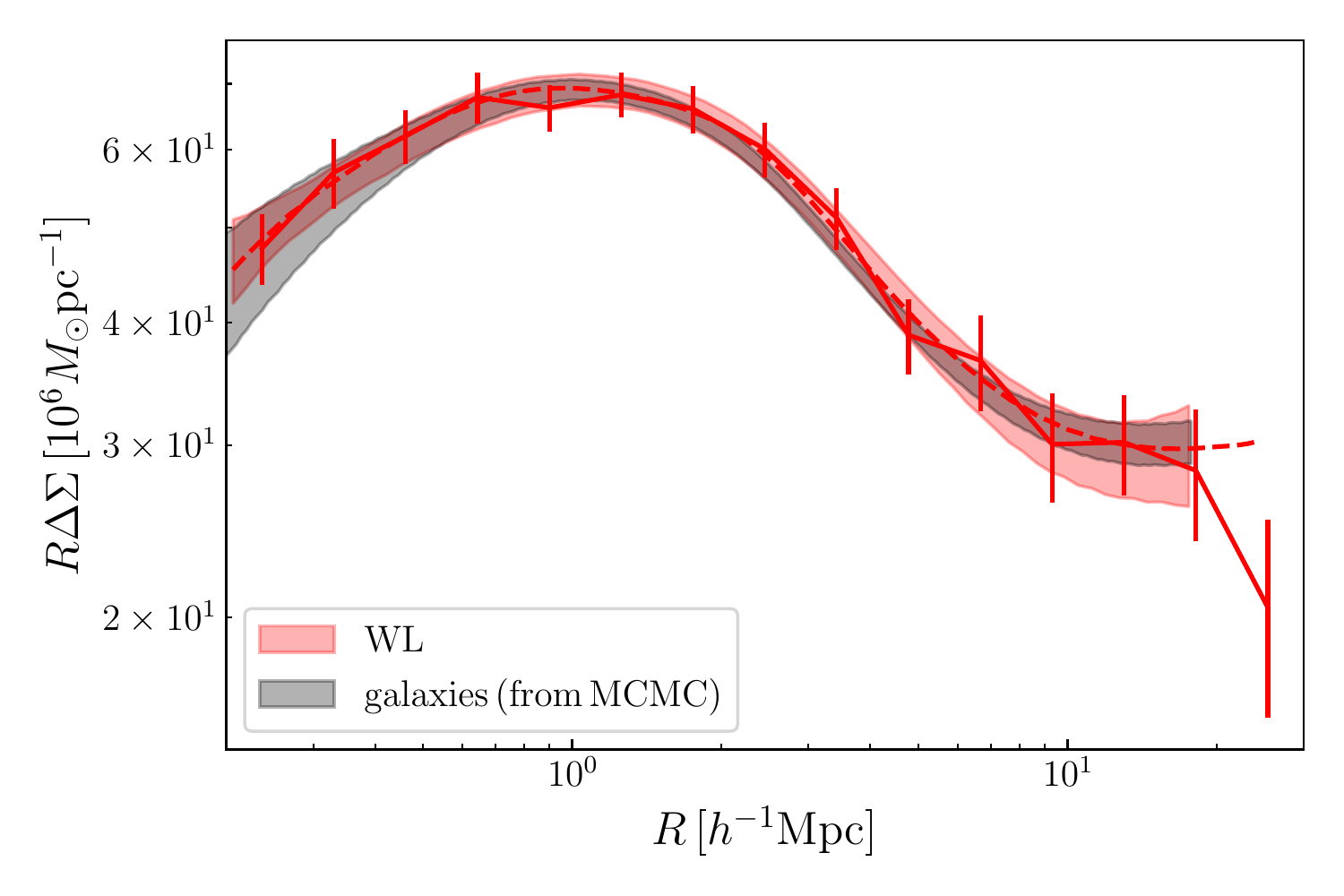}
\includegraphics[width=0.49\linewidth]{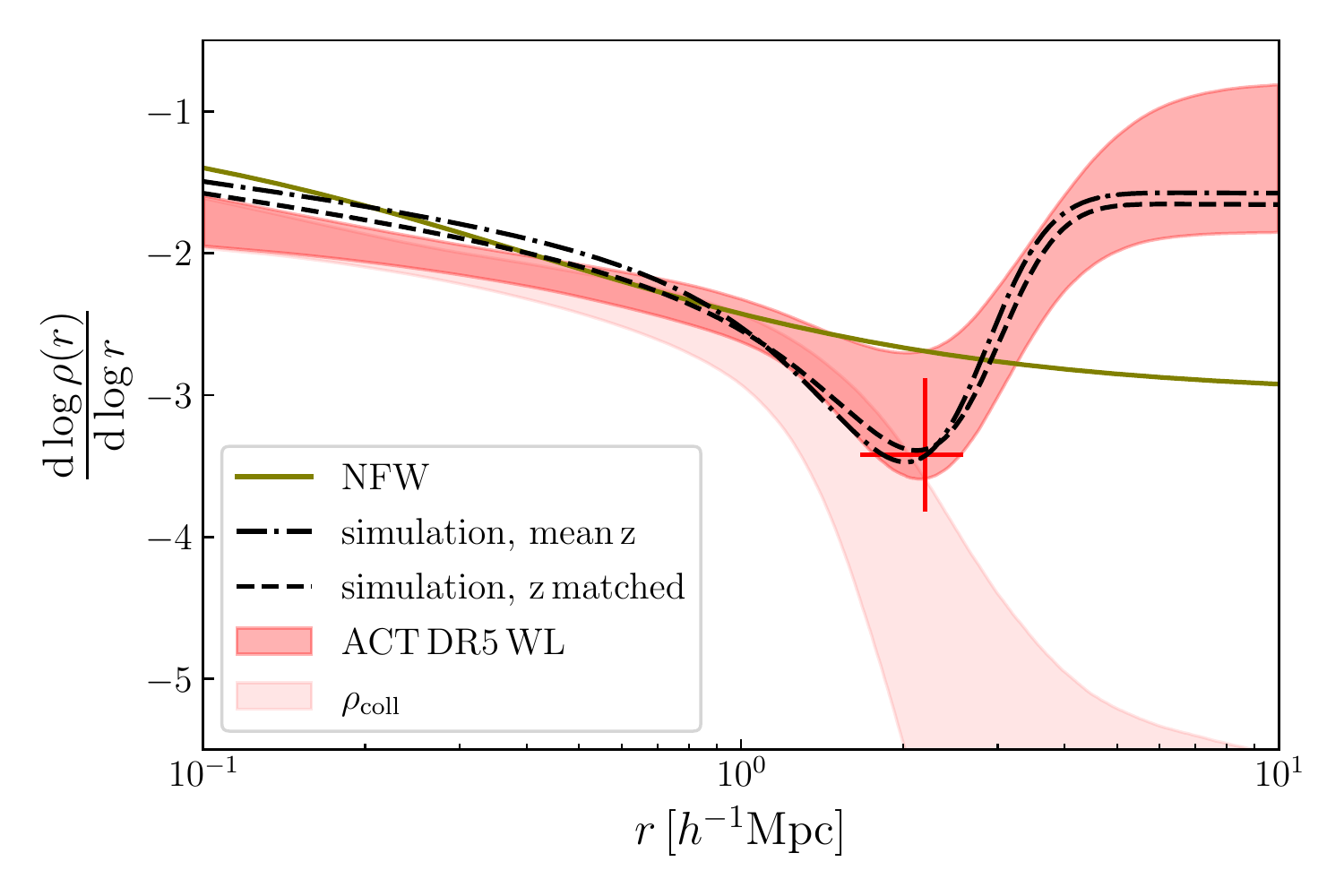}
\caption{
\textit{Left panel:} The measured $\Delta\Sigma$ profile of SZ-selected clusters is shown by the points in red with error bars, along with the 68\% confidence interval from  MCMC fitting of the DK14 model (red band). 
The dashed red curve is the best fit model.
The gray band shows the corresponding model fit for the galaxy profile (the measurements are shown in Fig. \ref{fig:galaxy-profile}), rescaled to match the amplitude of the WL measurement. 
\textit{Right panel:} The logarithmic slope of the 3D  matter profile from model fitting of the weak lensing measurements. The shaded band shows the 68\% confidence region, with  
the red cross being the 1-$\sigma$ constraints on the splashback radius, $r_{\rm sp}$ and the corresponding slope.
The profile of DM particles around mass-matched halos in the $N$-body simulation (Sec.~\ref{sec:data-sim}) is shown by the black curves. The dash-dotted line is the logarithmic slope profile for simulated clusters at the mean redshift of our sample and the dashed line is the profile of simulated clusters with the complete redshift weighting. The lighter shaded region shows the contribution of the ``collapsed'' (analogous to the 1-halo) term in the fit.
We also plot the theoretical NFW profiles for our mass and redshift values, calculated with the \textsc{Colossus} package \citep{Colossus}, as the olive green line.
}
\label{fig:slope-fit}
\end{figure*}

\subsection{Measurement of the WL profile}
\label{sec:measure-dsigma}

The tangential shear of a background galaxy around the center of a DM halo lens is given as,
\begin{equation}
    \gamma_{\rm t} = -\gamma_1 \cos{2\phi} - \gamma_2 \sin{2\phi},
\end{equation}
where $\gamma_1$ and $\gamma_2$ are the two shear components in a Cartesian coordinate system, and $\phi$ is the position angle of the source galaxy with respect to the x-axis of the system.

The tangential shear is then related to the 2D surface density profile of the halo as,
\begin{equation}
    \Delta\Sigma (R) = \bar{\gamma_t}(R) \, \Sigma_{\rm c} (z_{\rm l},z_{\rm s}),
\end{equation}
where $z_{\rm l}$ and $z_{\rm s}$ represent the redshifts of the lens (the DM halo) and the source,
$\bar{\gamma_t}(R)$ the mean tangential shear at the radius of $R$, and 
\begin{equation}
    \Sigma^{-1}_{\rm c} (z_{\rm l},z_{\rm s}) = \frac{4\pi G}{c^2} (1+ z_{\rm l}) \chi(z_{\rm l}) \left[ 1 - \frac{\chi(z_{\rm l})}{\chi(z_{\rm s})} \right]
\end{equation}
is the inverse critical density, with $\chi(z)$ representing the comoving distance to the redshift $z$ (see Eq.~\ref{eq:deltaSigma} for the definition of $\Delta\Sigma(R)$).

The cross component of the shear of a background galaxy is defined as:
\begin{equation}
    \gamma_\times = \gamma_1 \sin{2\phi} - \gamma_2 \cos{2\phi}.
\end{equation}
Note that, for any isotropic lens, $\Delta\Sigma_\times = \bar{\gamma}_\times \Sigma_{\rm c} (z_{\rm l}, z_{\rm s})$ is zero by angular symmetry.
We thus use the measurement of $\Delta\Sigma_\times$ as a null test.

We measure $\Delta\Sigma(R)$ using the estimator 
\begin{equation}
    \Delta\tilde{\Sigma} (R) = \frac{\sum_{ij} s^{ij} \gamma_{\rm t}^{ij} (R) }{\sum_{ij} s^{ij} \Sigma^{-1}_{\rm c,MC}(z^i_{\rm l},z^j_{\rm s}) (\mathcal{R}^j+\mathcal{R}_{\rm s})} ,
\end{equation}
where $i$ represents the lenses, $j$ the sources, $\mathcal{R}$ the shear response from \textsc{Metacalibration}, $\mathcal{R}_{\rm s}$ the selection response, and 
\begin{equation}
    s^{ij} = \omega^j \Sigma^{-1}_{\rm c,mean}(z^i_{\rm l},z^j_{\rm s})
\end{equation}
is the weight applied to optimize the measurement, where $\omega^j$ is the square inverse of the measured shear uncertainty of the $j$-th source (see Sec. 4.3 in \citet{Y3shape} for details). 
Note that the selection response terms do not apply to individual galaxies, but to the full sample of source galaxies.
Here, $\Sigma^{-1}_{\rm c,MC}$ represents the inverse critical density with the redshift of the source randomly chosen from the probability distribution given by the BPZ photo-$z$ estimation algorithm, and $\Sigma^{-1}_{\rm c,mean}$ that evaluated at the mean redshift from BPZ.
In addition, we exclude source galaxies whose photo-$z$'s lie within $\Delta z=0.1$ of the lens cluster, to reduce the contamination from the foreground galaxies to the source catalog. 
Then the measured $\Delta\Sigma(R)$ is related to the projected 2D density profile by Eq.~(\ref{eq:deltaSigma}).
See \citet{McClintock19} for a detailed validation of this estimator.

We stack the clusters and calculate $\Delta\tilde{\Sigma}(R)$ in 15 cluster-centric radial bins between 0.2 and 30 $h^{-1} {\rm Mpc}$, evenly spaced in log-space. 
Below $0.2h^{-1} {\rm Mpc}$, the crowding of  galaxies near the centers of the clusters hinders robust measurements of the background shears so that $\Delta\tilde{\Sigma}(R)$ becomes uncertain.
Therefore, we exclude the region below $0.2h^{-1} {\rm Mpc}$ from our WL measurement. 
Also, the DK14 model that we use was calibrated with simulations only up to $\sim 9 R_{\rm vir}$ (or $\sim$16$h^{-1} {\rm Mpc}$ for the mass and redshift of our clusters). We use this as the maximum scale in our model fitting for $\Sigma_g$; for the WL profile, we include one  additional data point (maximum $R=21.5 h^{-1} {\rm Mpc}$)  because $\Delta\Sigma(R)$ depends on the surface density at all radii smaller than $R$ so it is weighted towards smaller scales.

Also, to remove possible additive biases on the shear and reduce uncertainties on large scales, we calculate the $\Delta\tilde{\Sigma}(R)$ around the random points (20 times the number of the clusters) and subtract it from the signal around the clusters. We refer readers to Sec. 4.1.3 of \citet{McClintock19} and the references therein for detailed justification of it.
\citet{Wu19} show that statistical errors of cluster weak lensing at large scales are dramatically reduced by subtracting the profile around random points.

We use \texttt{treecorr} \citep{treecorr} to produce $\Delta\tilde{\Sigma}(R)$. The covariance matrix for the measurement is estimated using the jackknife method \citep{Norberg09}, with 100 patches having similar areas, which gives $\sim$45 square degrees per patch.

\subsection{Boost factor correction}
\label{sec:boost}

The photo-$z$ estimation for our galaxy sample comes with non-negligible errors.
Because of this error, the galaxies that are at or in front of the clusters would leak into our source sample and therefore dilute the WL signal.
Thus, in order to make a robust estimation of the surface density profile using WL, one must correct for this contamination in the source sample.
This is so-called ``boost factor'' correction, which we call $\mathcal{B}(R)$ and which we use to multiply our $\Delta\Sigma$ estimator:
\begin{equation}
    \Delta\tilde{\Sigma}_{\rm corr} = \mathcal{B}\Delta\tilde{\Sigma}.
\end{equation}

There are two methods generally used.
In the first method \citep{Sheldon2004,Mandelbaum2008}, weighted counts of the source galaxies around the lens sample and around random positions are calculated and one takes the fraction between them:
\begin{equation}
\label{Eq:boost_count}
    \mathcal{B}(R) = \frac{N_{\rm rand}}{N_{\rm lens}}\frac{\sum_{ij} s^{ij}}{\sum_{kj} s^{kj}},
\end{equation}
where $N_{\rm lens}$ and $N_{\rm rand}$ represent the number of  lenses and  random points, respectively, and $i$, $j$ and $k$ run over lenses, source galaxies and random points, respectively.

In the second method \citep{Varga19}, one decomposes the probability distribution of the source galaxy redshifts into two parts: the contamination part and the true source distribution part:
\begin{equation}
    P(z|R) = f_{\rm cl}(R) \, P_{\rm cont}(z|R) + (1-f_{\rm cl}(R)) \, P_{\rm bg}(z),
\end{equation}
where $f_{\rm cl}(R)$ is the fraction of the contamination as a function of radius, $P_{\rm cont}(z|R)$ the probability distribution of the contaminating galaxies at  radius $R$, $P_{\rm bg}(z)$ the probability distribution of the true background source sample.
Here, $P_{\rm bg}(z)$ is calculated around the random points and we assume a Gaussian distribution for $P_{\rm cont}(z|R)$ (see Sec. 3.2.5 of \citet{Varga19} for the validation of the Gaussian assumption).
Therefore, the free parameters in this method are $f_{\rm cl}(R)$ and the width of $P_{\rm cont}(z|R)$ for which we can find the best fit value, given $P_{\rm bg}(z)$ and $P(z|R)$ from the data. The boost factor is related to $f_{\rm cl}$ as,
\begin{equation}
    \mathcal{B}(R) = \frac{1}{1-f_{\rm cl}(R)}.
\end{equation}

In \citet{Varga19}, it is shown that the $P(z)$ decomposition method correctly retrieves the true values of the boost factor, whereas the counting-based method (Eq.~\ref{Eq:boost_count}) tends to underestimate the boost factor. 
Hence, in this paper we use the  $P(z)$ decomposition method as our fiducial choice.
We have checked that the choice of the boost factor does not alter our results as it makes only a small difference to the inner profile. 
The covariance matrix of the boost factor is estimated via the jackknife method using the same configuration as in the previous section.

\begin{table}
	\centering
	\begin{tabular}{lll}
		Parameter & Prior & description \\ \hline
		$\log \rho_{\rm s}$ & $[-\infty,\infty]$ &  amplitude of the Einasto profile (Eq.3) \\
        $\log \alpha$ & $\mathcal{N}(\log(0.22),0.6^2)$ & parameter of the Einasto profile\\
        $\log r_{\rm s}$ & $[\log(0.1),\log(5.0)]$ & scale radius of the Einasto profile\\
		$\log r_{\rm t}$ & $[\log(0.5),\log(5.0)]$ & scale radius of $f_{\rm trans}$\\
		$\log \beta$ & $\mathcal{N}(\log(6.0),0.2^2)$ & first slope parameter of $f_{\rm trans}$\\
		$\log \gamma$ & $\mathcal{N}(\log(4.0),0.2^2)$ & second slope parameter of $f_{\rm trans}$\\
		$\log \rho_{\rm 0}$ & $[-\infty,\infty]$ &  amplitude of $\rho_{\rm infall}$\\
        $s_{\rm e}$ & [0.1,10.0] & log-slope of $\rho_{\rm infall}$\\
        $\ln c_{\rm mis}$ & $\mathcal{N}(-1.13,0.22^2)$ & miscentering amplitude\\
        $f_{\rm mis}$ & $\mathcal{N}(0.22,0.11^2)$ & miscentering fraction\\
        $\log \mu$ & $[-\infty,\infty]$ & inner mass contribution (WL)
	\end{tabular}
    \caption{Prior range of each model parameter. $\mathcal{N}(m,\sigma^2)$ represents a Gaussian prior with  mean $m$ and  standard deviation $\sigma$ (see Sec.~\ref{sec:model} and ~\ref{sec:model-fitting}). $r_{\rm s}$ and $r_{\rm t}$ are in a unit of $h^{-1} {\rm Mpc}$}
\label{tab:modeling_parameters}
\end{table}

\subsection{Measurement of the projected galaxy number density profile}
\label{sec:measure-sigmag}

We follow the method in \citet{Chang2017} and \citet{Shin19} to measure the projected galaxy   profile around clusters in our sample.
We first cross-correlate the ACT DR5 cluster sample (Sec.~\ref{sec:data-szcl}) with the DES Y3 galaxy sample (Sec.~\ref{sec:data-desgold}) in narrow redshift bins of $\Delta z=0.025$ using the Landy-Szalay estimator \citep{Landy93}. 
We apply this redshift binning only to the cluster sample assuming they are located at the mid-point of the corresponding bin.
We have checked that this approximation does not change the measured data points significantly, given the level of the uncertainty of the data.
To avoid the uncertainty of the photo-$z$ estimation, we assume that all galaxies are located at the cluster redshift and apply an additional cut on the absolute magnitude, $M_i<-19.87$, which corresponds to the apparent magnitude cut, $m_i<22.5$, at the maximum redshift of 0.7 to ensure the same maximum luminosity of galaxies over the whole range of redshift $(0.15<z<0.7)$. 
The correlation function then automatically selects the galaxies that are physically correlated with the clusters, while uncorrelated galaxies (at different redshifts) are not reflected in the correlation.
In this way, we avoid the systematic errors induced by the uncertainty of the galaxy photo-$z$ estimates.

To obtain the mean correlation function over all redshift, $\omega(R)$, the computed correlation functions for each redshift bin, $\omega(R,z_i)$, are averaged with the number of clusters in each redshift bin as weights:
\begin{equation}
    \omega(R) = \frac{\sum_i N_{{\rm cl},i} \omega(R,z_i)}{\sum_i N_{{\rm cl},i}},
\end{equation}
where $N_{{\rm cl},i}$ is the number of clusters in the $i$-th redshift bin.
This $\omega(R)$ is related to the average mean-subtracted projected galaxy  profile around the cluster sample as:
\begin{equation}\label{eq:sigmag_data}
    \Sigma_{\rm g}(R) = \bar{\Sigma}_{\rm g} \, \omega(R),
\end{equation}
where $\bar{\Sigma}_{\rm g}$ represents the average surface number density of the galaxy sample:
\begin{equation}\label{eq:sigmagbar_data}
    \bar{\Sigma}_{\rm g} = \frac{\sum_i N_{{\rm cl},i} \bar{\Sigma}_{{\rm g},i}}{\sum_i N_{{\rm cl},i}},
\end{equation}
with $\bar{\Sigma}_{{\rm g},i}$ being the average surface number density of the galaxies used in each redshift bin.

$\Sigma_{\rm g}(R)$ is calculated in 25 radial bins between 0.2 and 60 $h^{-1} {\rm Mpc}$, evenly spaced in comoving log-space, using \texttt{treecorr} \citep{treecorr}.
Similar to the scale cut for lensing in Section~\ref{sec:measure-dsigma}, we use  radial bins larger than $\sim0.2h^{-1} {\rm Mpc}$, since below that scale the BCG and intracluster light and the crowding of  galaxies may interfere with  galaxy detection.
We also exclude  bins larger than $\sim$16$h^{-1} {\rm Mpc}$ from the fitting since they lie above $9r_{\rm vir}$, which is  the radial range over which the theoretical model of \citet{Diemer2014} was calibrated with simulations.
The covariance matrix of $\Sigma_{\rm g}(R)$ is estimated using the jackknife method \citep{Norberg09}, with 100 patches of similar size as before.
We have generated data points with different numbers of jackknife patches (50 and 150) and checked that our data points are stable up to $\sim 40 h^{-1} {\rm Mpc}$.

\begin{table*}
	\centering
    \bgroup
    \def\arraystretch{1.0}
    \setlength{\tabcolsep}{4pt}
	\begin{tabular}{cccccccccccc}
		Sample & $\log \alpha$ & $\log (r_{\rm s})$ & $\log r_{\rm t}$ & $\log \beta$ & $\log \gamma$ & $s_{\rm e}$ & $f_{\rm mis}$ & $\ln c_{\rm mis}$ & $r_{\rm sp}$ [$h^{-1}$Mpc] & $\frac{d\log\rho}{d\log r}(r_{\rm sp})$ & $\frac{d\log\rho_{\rm coll}}{d\log r}(r_{\rm sp})$ \\ \hline
		fiducial ($\Delta\Sigma$) & $-0.91^{+0.27}_{-0.26}$ & $-0.88^{+0.38}_{-0.03}$ & $0.34^{+0.17}_{-0.14}$ & $0.76^{+0.19}_{-0.22}$ & $0.69^{+0.08}_{-0.32}$ & $1.60^{+0.25}_{-0.80}$ & $0.20^{+0.07}_{-0.11}$ & $-1.22^{+0.30}_{-0.18}$ & $2.20^{+0.39}_{-0.54}$ & $-3.42^{+0.54}_{-0.40}$ & 
		$-5.20^{+1.27}_{-0.61}$\\
		fiducial ($\Sigma_{\rm g})$ & $-0.68^{+0.10}_{-0.25}$ & $-0.65^{+0.11}_{-0.20}$ & $0.32^{+0.08}_{-0.08}$ & $0.81^{+0.12}_{-0.25}$ & $0.67^{+0.16}_{-0.22}$ & $1.59^{+0.07}_{-0.10}$ & $0.18^{+0.07}_{-0.08}$ & $-1.13^{+0.18}_{-0.25}$ & $2.07^{+0.12}_{-0.26}$ & $-3.40^{+0.32}_{-0.17}$ & $-5.50^{+1.15}_{-0.46}$\\
        high lum & $-0.67^{+0.12}_{-0.25}$ & $-0.72^{+0.15}_{-0.18}$ & $0.29^{+0.16}_{-0.10}$ & $0.61^{+0.32}_{-0.09}$ & $0.71^{+0.08}_{-0.31}$ & $1.58^{+0.10}_{-0.11}$ & $0.17^{+0.07}_{-0.08}$ & $-1.16^{+0.26}_{-0.19}$ & $1.83^{+0.28}_{-0.27}$ & $-3.26^{+0.34}_{-0.17}$ & $-4.89^{+0.96}_{-0.70}$\\
		low lum & $-0.82^{+0.18}_{-0.17}$ & $-0.76^{+0.20}_{-0.14}$ & $0.33^{+0.11}_{-0.07}$ & $0.82^{+0.14}_{-0.25}$ & $0.65^{+0.16}_{-0.22}$ & $1.55^{+0.08}_{-0.12}$ & $0.20^{+0.06}_{-0.10}$ & $-1.19^{+0.22}_{-0.21}$ & $2.20^{+0.20}_{-0.26}$ & $-3.45^{+0.39}_{-0.18}$ & $-5.43^{+1.18}_{-0.49}$ \\
        high mass & $-0.78^{+0.12}_{-0.23}$ & $-0.69^{+0.12}_{-0.20}$ & $0.33^{+0.10}_{-0.08}$ & $0.80^{+0.15}_{-0.21}$ & $0.65^{+0.16}_{-0.22}$ & $1.56^{+0.11}_{-0.12}$ & $0.13^{+0.11}_{-0.06}$ & $-1.22^{+0.29}_{-0.31}$ & $2.15^{+0.18}_{-0.30}$ & $-3.29^{+0.34}_{-0.20}$ & $-5.31^{+1.09}_{-0.51}$ \\
        low mass & $-0.65^{+0.15}_{-0.40}$ & $-0.71^{+0.22}_{-0.16}$ & $0.29^{+0.13}_{-0.15}$ & $0.85^{+0.09}_{-0.38}$ & $0.66^{+0.13}_{-0.29}$ & $1.60^{+0.14}_{-0.13}$ & $0.27^{+0.05}_{-0.13}$ & $-1.16^{+0.18}_{-0.18}$ & $1.97^{+0.09}_{-0.52}$ & $-3.46^{+0.45}_{-0.12}$ & $-5.65^{+1.77}_{-0.18}$ \\
        high $z$ & $-0.71^{+0.18}_{-0.31}$ & $-0.60^{+0.13}_{-0.23}$ & $0.30^{+0.10}_{-0.07}$ & $0.86^{+0.14}_{-0.23}$ & $0.68^{+0.16}_{-0.21}$ & $1.41^{+0.08}_{-0.14}$ & $0.31^{+0.09}_{-0.15}$ & $-1.48^{+0.21}_{-0.12}$ & $2.06^{+0.18}_{-0.24}$ & $-3.71^{+0.50}_{-0.30}$ & $-5.76^{+1.28}_{-0.62}$\\
        low $z$ & $-1.11^{+0.52}_{-0.22}$ & $-1.00^{+0.34}_{-0.06}$ & $0.28^{+0.29}_{-0.05}$ & $0.78^{+0.14}_{-0.30}$ & $0.55^{+0.20}_{-0.20}$ & $1.76^{+0.24}_{-0.11}$ & $0.11^{+0.14}_{-0.02}$ & $-1.14^{+0.27}_{-0.11}$ & $1.76^{+0.57}_{-0.26}$ & $-2.95^{+0.25}_{-0.19}$ & $-4.10^{+0.56}_{-1.33}$
	\end{tabular}
    \egroup
    \caption{1$\sigma$ ranges of the best-fit parameters in different samples, including the model parameters (Sec.~\ref{sec:model}), splashback location ($r_{\rm sp}$) and the minimum logarithmic slope at $r_{\rm sp}$. The last column gives the 1$\sigma$ range of the logarithmic derivative of $\rho_{\rm coll}$, which provides a measure of the deviation from the (no splashback) NFW profile.
    The mean mass and the redshift of the fiducial sample are $M_{\rm 500c} = 2.7\times10^{14}h^{-1}M_{\odot}$ and $z=0.455$, respectively.
    `high lum' represents the clusters correlated with galaxies with $M_i<-20.9$, `low lum' those correlated with galaxies with $-20.9<M_i<-19.9$; `high mass' those with $\langle M_{\rm 500c}\rangle = 3.75\times10^{14}h^{-1}M_{\odot}$, `low mass' those with $\langle M_{\rm 500c} \rangle = 2.14\times10^{14}h^{-1}M_{\odot}$; `high $z$' those with redshift between 0.45 and 0.70, and `low $z$' those with redshift between 0.15 and 0.45 (see Appendix~\ref{app:split} and \ref{app:galaxy_dimmer}).
    Note that we do not show the parameters $\rho_0$ and $\rho_{\rm s}$, since they do not contain information related to $r_{\rm sp}$.
    The 1-$\sigma$ constraint on the parameter $\mu$ in  $\Delta\Sigma$ (Sec.~\ref{sec:model-fitting}, Table~\ref{tab:modeling_parameters}) is $1.09^{+0.01}_{-0.02}$. $r_{\rm s}$ and $r_{\rm t}$ are in a unit of $h^{-1} {\rm Mpc}$.
    See Appendix~\ref{app:prior_test} for the effects of  miscentering and the priors on the other model parameters.
    }
    \label{tab:results}
\end{table*}

\subsection{Model fitting}
\label{sec:model-fitting}

We adopt a Gaussian likelihood for the profile measurements:
\begin{equation}
    \ln \mathcal{L}[\vec{d} | \vec{m}(\vec{\theta}) ] = -\frac{1}{2}\left[\vec{d} - \vec{m}(\vec{\theta})\right]^{\rm T} \mathcal{C}^{-1} \left[\vec{d} - \vec{m}(\vec{\theta}) \right],
\end{equation}
where $\vec{d}$ represents the data vector ($\Delta\Sigma$ or $\Sigma_{\rm g}$), $\vec{m}(\vec{\theta})$ the model evaluated at parameters $\vec{\theta}$, and $\mathcal{C}$ is the covariance estimated using the jackknife resampling method.  The posterior on the model parameters is then
\begin{equation}
    \ln \mathcal{P}(\vec{\theta} | \vec{d}) = \ln \big[ \mathcal{L}(\vec{d} | \vec{m}(\vec{\theta}) ){\rm Pr}(\vec{\theta}) \big],
\end{equation}
where ${\rm Pr}(\vec{\theta})$ are the priors applied on $\vec{\theta}$. 

Previous analyses \citep{Hilton_2020} have indirectly used weak lensing to calibrate the mass-observable relation for the ACT clusters; for all the clusters, the WL correction factor of $1/(0.71 \pm 0.07)$\footnote{The ratio of the mean mass derived from the mass-richness relation of the DES \textsc{redMaPPer} clusters \citep{McClintock19}, to that from the SZ mass-observable relation} is applied to the mass from the SZ mass-observable relation.
The average uncertainty on the mass, $M_{500c}$, from these measurements is $\sim$23\% which includes the statistical uncertainty as well as that from the WL correction factor. 
To be conservative, we assume the uncertainties of the individual masses are 100\% correlated.
We then use this mass information as a prior on our model fitting. 
We apply a Gaussian prior on the total mass of the profile, $M_{\rm 500c}=2.72 \pm 0.68 \times 10^{14} h^{-1} M_{\odot}$ (25\% uncertainty). 
The $R_{\rm 500c}$ is calculated at the mean redshift of the clusters (0.455) with the mass-concentration relation from \citet{Diemer2019}, which gives $R_{\rm 500c}=0.96h^{-1} {\rm Mpc}$ for our mass. 

We adopt priors on the model parameters (Sec.~\ref{sec:model}) that are similar to those of \citet{Chang2017} and \citet{Shin19}.  The applied priors are listed in Table~\ref{tab:modeling_parameters}.  
The only change with respect to \citet{Shin19} is that the minimum of $r_{\rm t}$ is $0.5 h^{-1} {\rm Mpc}$, since we are certain that for clusters with our mass the transition between the 1-halo and the infall regime happens above that radius. 
Our adopted priors on the miscentering parameters are identical to those for \textsc{redMaPPer} clusters \citep{Rykoff2016}, since we adopt the BCG locations measured by \textsc{redMaPPer} as the cluster center.
For those that do not have the \textsc{redMaPPer} counterparts, we use SZ centers. 
However, the fraction of those without the \textsc{redMaPPer} counterparts is negligible, so that it does not affect our fitting.
We refer the reader to Appendix~\ref{app:prior_test} for tests on the effect of miscentering and the priors on the model paramters.

We sample the model posterior using the affine invariant Markov Chain Monte Carlo method introduced by \cite{mcmc} and implemented in the \texttt{emcee} package \citep{emcee13}. 
Note that the number of parameters is 10 for $\Sigma_{\rm g}$ and 11 for $\Delta\Sigma$, while  the number of data points is 20 for $\Sigma_{\rm g}$ and 14 for $\Delta\Sigma$. Thus for the lensing fits, the number of parameters is approaching the number of data points. 
One of our main goals is to use model fits that smoothly approximate the measured data points and allow us to estimate the correct logarithmic slope profiles. This exercise is valid and useful even for a large number of model parameters. 
The convergence of the MCMC chains is confirmed by splitting our chains into 5 pieces and comparing the results. 

\section{Results}
\label{sec:result}

\subsection{Mass profiles from lensing}
\label{sec:result-WL}

We begin by presenting the results of our lensing measurements and the corresponding profile fits.
The top panel of Fig.~\ref{fig:measurement-wl} shows  the measured cross component of the lensing signal  ($\Delta\Sigma_{\times}$) for our  sample of clusters, which serves as a null test of our lensing measurements.  This measurement is consistent with zero as expected: null-$\chi^2 / {\rm dof} = 14.7 / 15$, where 15 is the number of data points. 
The boost factors from the two different calculation methods (Sec.~\ref{sec:boost}) are shown in the bottom panel of Fig.~\ref{fig:measurement-wl}.
The boost factor from the $P(z)$ decomposition method shows a somewhat smaller value in the first radial bin, although the difference is not statistically significant and does not affect our fitting results.  As argued in Sec.~\ref{sec:boost}, the $P(z)$ decomposition method results in a better estimate of the boost factor correction, so we adopt it as our fiducial choice when fitting the profiles.

In the left panel of Fig.~\ref{fig:slope-fit}, we show measured $\Delta\Sigma$ (Sec.~\ref{sec:measure-dsigma}) profile around our cluster sample. 
With the $\Delta\Sigma$, the boost factor and our halo model (Sec.~\ref{sec:model}), we use the MCMC methods described previously (Sec.~\ref{sec:model-fitting}) to sample from the posterior on the model parameters. 
The red shaded region in the left panel of Fig.~\ref{fig:slope-fit} represents the 1-$\sigma$ range of our posterior on $\Delta\Sigma$.  The minimum $\chi^2$ is 1.95 with 3 degrees of freedom (14 data points with 11 fitting parameters).
The full results for the MCMC-fitted model parameters are listed in Table~\ref{tab:results} including the results from the sample split tests described in Appendix~\ref{app:split} and \ref{app:galaxy_dimmer}.

We then calculate the 3D logarithmic slope ($d \log \rho(r) / d \log r$) of the matter profile using the constrained halo model parameters from the MCMC chain and identify the splashback radius as the location of the minimum slope.
In the right panel of Fig.~\ref{fig:slope-fit}, the shaded region represents the 68\% credible interval for the 3D logarithmic slope of the matter density profile inferred from $\Delta\Sigma$.
The red cross represents the 68\% credible interval for the splashback radius, $r_{\rm sp}$, and the corresponding slope at $r_{\rm sp}$.
The 1-$\sigma$ constraint on $r_{\rm sp}$ from WL is $2.20^{+0.39}_{-0.54} h^{-1}{\rm Mpc}$, and that of the logarithmic slope at $r_{\rm sp}$ is $-3.42^{+0.54}_{-0.40}$.

Also shown in the same figure with the black dash-dotted line is the 3D logarithmic slope of the DM particle profile around the halos from the MDPL2 $N$-body simulation at the mean redshift of our sample.
In addition, the DM profile from a lower resolution simulation for which we match the redshift distribution to our cluster sample is plotted as a black dashed line (see Sec.~\ref{sec:data-sim}).
The splashback radii of the simulation profiles are $r_{\rm sp}=2.03 h^{-1}{\rm Mpc}$ at the mean redshift, and $r_{\rm sp}=2.12 h^{-1}{\rm Mpc}$ for the redshift-matched simulation halos. 
As can be seen in the figure, the splashback radius from the WL profile agrees with that from the $N$-body simulations well within 1-$\sigma$. 
Also, plotted as the olive green line is the theoretical prediction of the NFW profile having the same mean mass and the redshift as ours. 
One can see that the logarithmic slope of the $\rho_{\rm coll}$ (1-halo term, shaded in light red) is steeper by over $\sim$2.5$\sigma$ (99.7\% of the posterior) than the slope ($\sim -2.7$) of the NFW profile at $r_{\rm sp}$, which can be taken as evidence for a splashback-like truncation of the density profile \citep{Baxter17}.

\begin{figure*}
\centering
\includegraphics[width=0.49\linewidth]{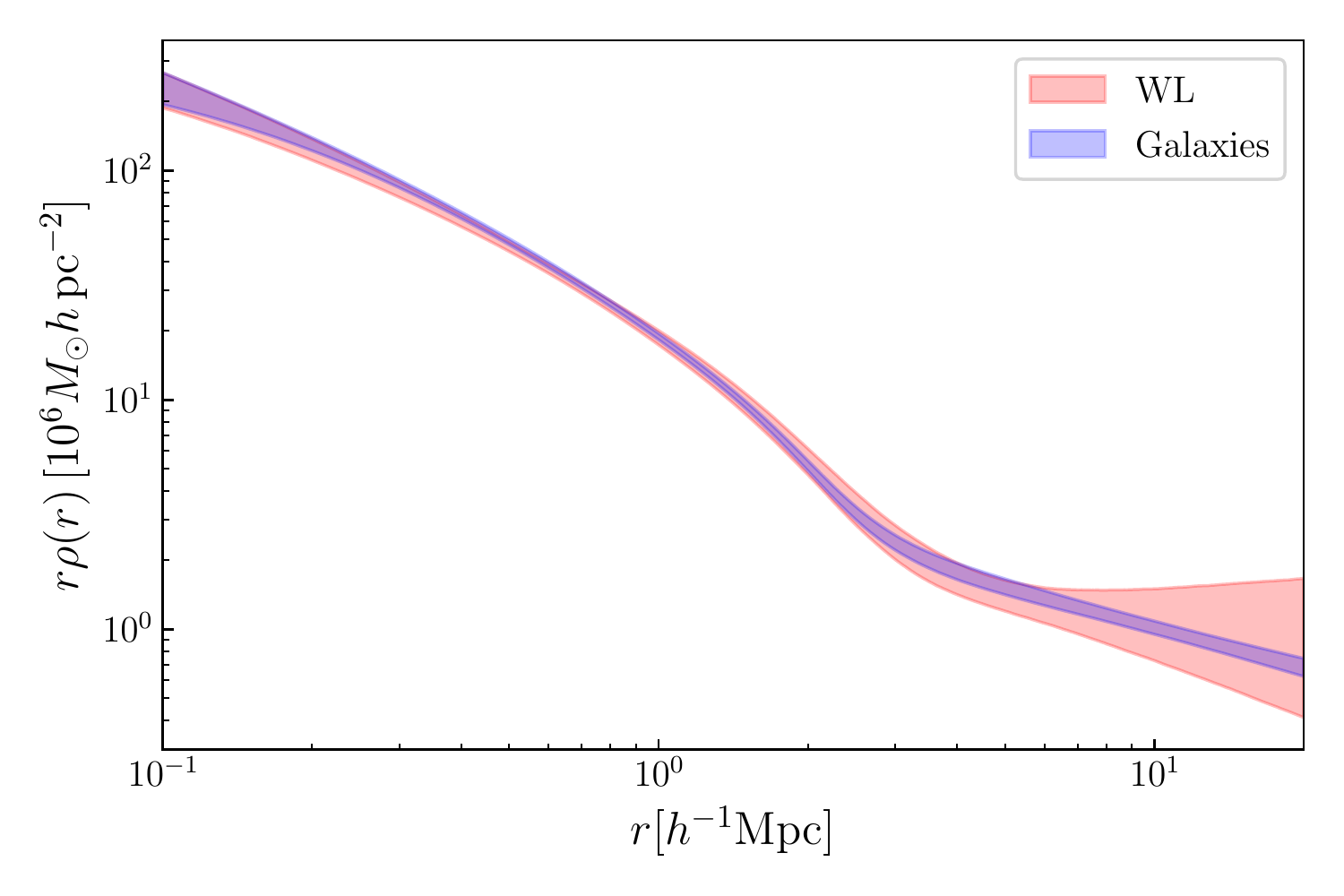}
\includegraphics[width=0.49\linewidth]{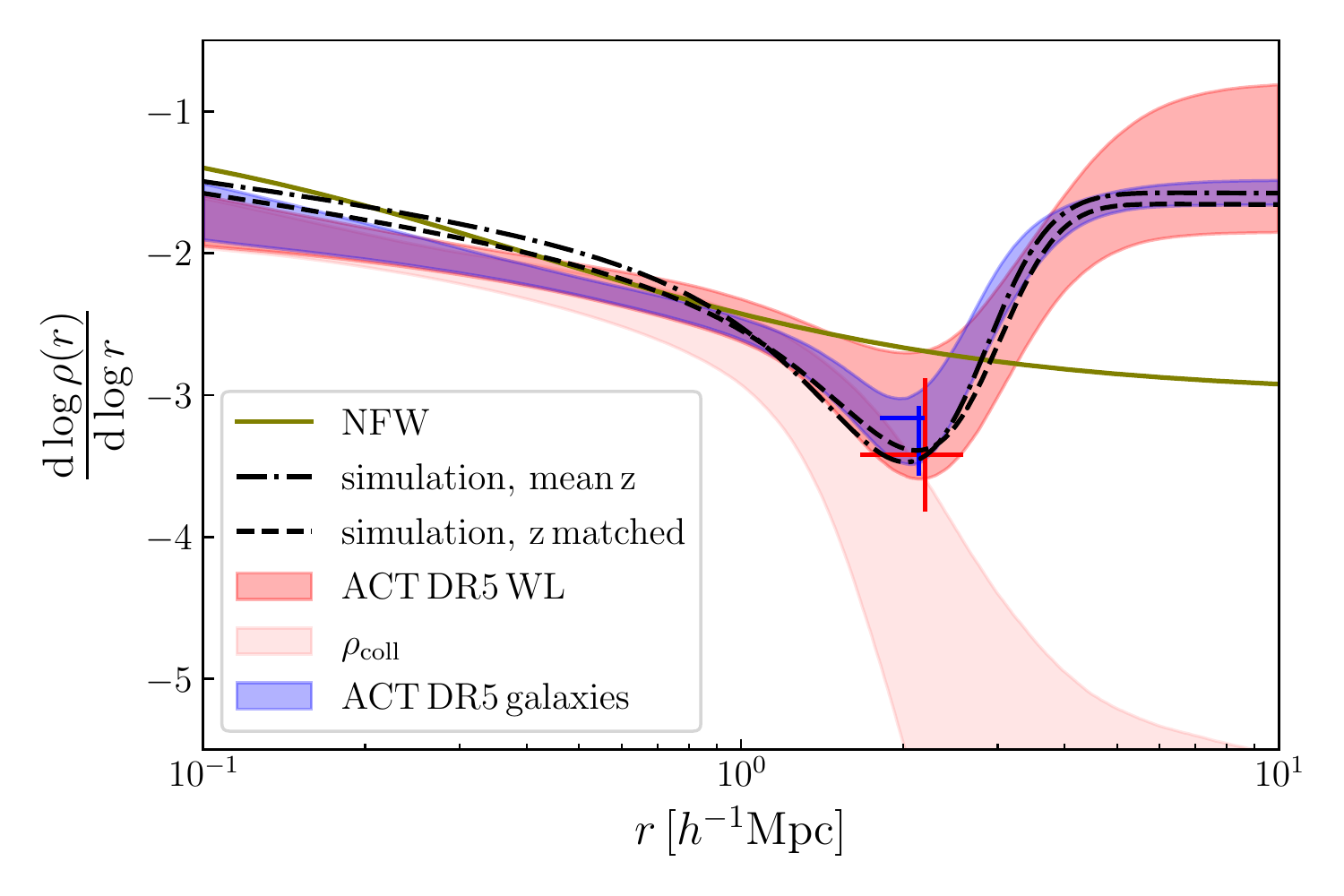}
\caption{\textit{Left panel:} The 3D density profiles inferred from the measured  WL (red) and projected galaxy number density profiles (blue) as described in  Sec.~\ref{sec:model-fitting}. 
The galaxy density profile normalization is shifted so that the difference between the two curves is minimized. 
\textit{Right panel:} The  logarithmic slope of the 3D dark matter profile inferred from the WL (red shaded region) and the galaxy density  (blue shaded region) profiles via the model fits as in the previous figures. 
The crosses represent the 1-$\sigma$ constraints on the splashback radius, $r_{\rm sp}$, and the corresponding slopes. The mean dark matter profiles from  mass-matched halos from $N$-body simulations (Sec.~\ref{sec:data-sim}) are  shown as the black lines. See Sec. \ref{sec:result-WL}, \ref{sec:result_comp} and \ref{sec:hydro_comp} for details. 
}
\label{fig:gal_wl_comp}
\end{figure*}

\begin{figure*}
\centering
\includegraphics[width=0.49\linewidth]{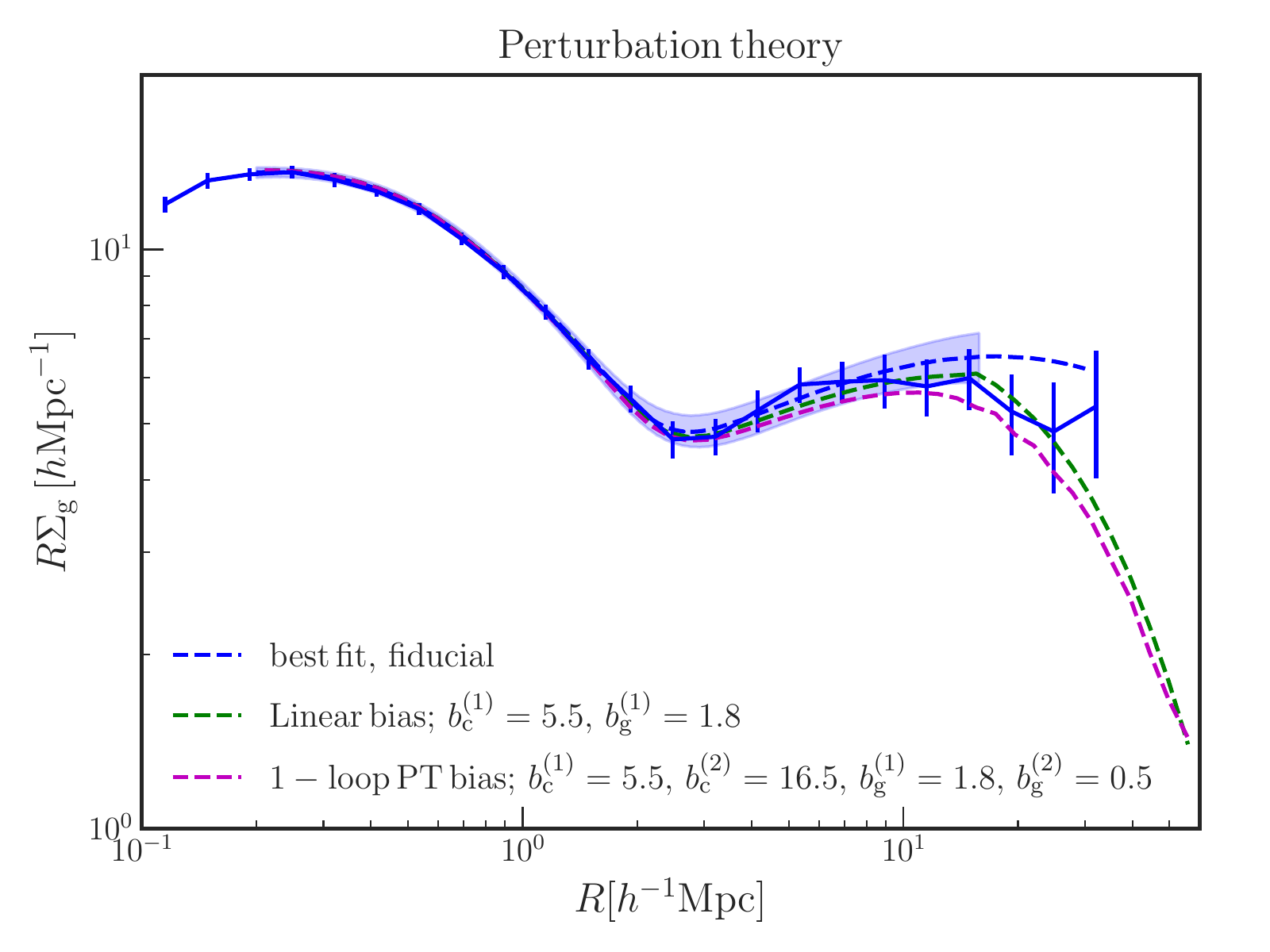}
\includegraphics[width=0.49\linewidth]{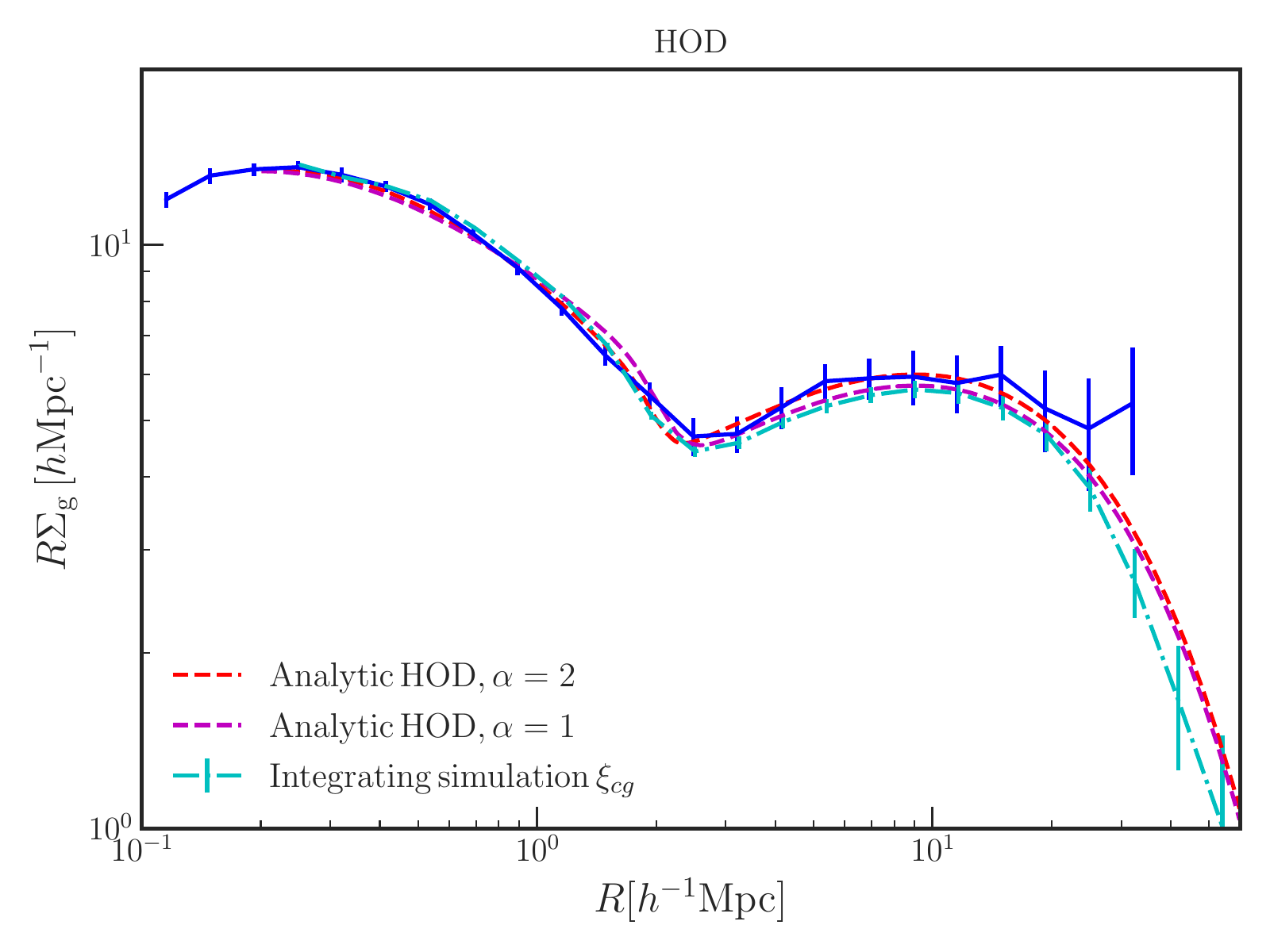}
\caption{
\textit{Left panel:} the projected galaxy number density profile around clusters is shown by the blue points with error bars. The blue dashed line and the shaded region represent the best fit DK14 model and the 68\% confidence  interval   (see Sec.~\ref{sec:model-fitting}).  The green  (purple) dashed curves are the result of patching the DK14 prediction with linear (second order perturbation theory) galaxy bias predictions at 15$h^{-1} {\rm Mpc}$. The legend shows the bias parameters used in the two models (see Sec. \ref{sec:theory_comp}). 
\textit{Right panel:} predictions from the standard halo model with  a Halo Occupation Distribution approach for the galaxy distribution. The dashed curves show the prediction from the analytical halo model  and from N-body simulations mocks generated at the best-fit parameters from the analytical model fit (cyan with error bars). The HOD based analytic predictions have some freedom in the transition region around the splashback radius, parameterized by $\alpha$ (Eq. 36), but typically predict scale dependent differences between the galaxy and mass profiles that are not seen in the data.  
}
\label{fig:galaxy-profile}
\end{figure*}

\subsection{Galaxy number density profiles}
\label{sec:result-gal}

In the left panel of Fig.~\ref{fig:galaxy-profile}, the blue data points with error bars show the measured galaxy profile (Sec.~\ref{sec:measure-sigmag}).
The 1-$\sigma$ range from MCMC model fitting (best fit $\chi^2$/$\rm dof$=2.63/7) is over-plotted as the blue shaded region.
Note that the fit is reasonable up to about $9r_{\rm vir} \sim 16 h^{-1} {\rm Mpc}$, which is the regime of validity of the DK14  model (see Sec.~\ref{sec:model} and Sec.~\ref{sec:measure-sigmag}).
The best fit curve (blue dashed) deviates from the measurements  above $\sim$16$h^{-1} {\rm Mpc}$ as shown above for the WL profile (Sec.~\ref{sec:result-WL}).
We discuss the implication of this finding in Sec.~\ref{sec:theory_comp} where other models to describe cluster-galaxy correlations are presented. 

\subsection{Comparison between mass and galaxy profiles}
\label{sec:result_comp}

We now turn to a comparison of the lensing and galaxy profiles. The left panel of Fig. \ref{fig:gal_wl_comp} shows the 3D profiles of the galaxy and total matter density obtained from the best-fit models. 
The $y$-axis for the galaxy density in this figure has been rescaled to minimize the difference between the matter and galaxy profiles (the scaling factor is 5.16).
The shapes of the profiles show remarkable agreement over all scales shown in the figure.   
The right panel of Fig.~\ref{fig:gal_wl_comp} compares the 3D logarithmic slope of the WL profiles (red), and that of the projected galaxy number density profile (blue). The crosses in the figure represent the 1-$\sigma$ ranges of the splashback location for the galaxy density profile (blue) and the WL profile (red).
The fit gives $r_{\rm sp}$ of the  galaxy profile: $2.07^{+0.12}_{-0.26} h^{-1} {\rm Mpc}$.
It is in good agreement with that of the WL profile which gives $2.20^{+0.39}_{-0.54} h^{-1} {\rm Mpc}$.
The logarithmic slope at the splashback radius is also in agreement: for galaxies the slope at $r_{\rm sp}$ is $-3.40^{+0.32}_{-0.17}$ and for weak lensing it is $-3.42^{+0.54}_{-0.40}$. The latter is well within the range of slopes expected for the cluster mass halos from simulations and corresponds to peak height $\nu\approx 2.5$\footnote{$\nu = \frac{\delta_{\rm c}}{\sigma(M,z)}$, where $\delta_{\rm c} = 1.686$ is the critical overdensity for collapse and $\sigma(M,z)$ the rms density fluctuation in a sphere of radius $R$. See e.g. Sec. 2.3 of \citet{Diemer2014} for details.} and accreting mass at a moderate rate of $1<\Gamma<2$, where $\Gamma=\Delta \log (M_{\rm vir}) / \Delta \log (a)$ (see Figs 5 and 10 in DK14). 

The agreement between the galaxy number density profile and the WL profile indicates that our galaxy sample ($M_i<-19.87$) closely follows the underlying matter distribution for our fiducial cluster sample. 
This agreement was anticipated in the more direct comparison  in the left panel of Fig.~\ref{fig:slope-fit}.

We  discuss the implication of these findings in Sec.~\ref{sec:discussion}.
We also explore how a different luminosity selection for the galaxies, along with other data splits,  might affect our results in Appendix \ref{app:split}. Next we compare model predictions for the galaxy profile with the measurement from data.

\subsection{Models of cluster-galaxy correlations}
\label{sec:theory_comp}

We present exploratory comparisons of our measurements and predictions based on different theoretical approaches. First, we match the DK14 model with perturbation theory (PT) on large scales. We call this approach Hybrid PT as it uses PT for the galaxy and cluster bias parameters and the nonlinear matter correlation function calibrated from N-body simulations. 

\subsubsection{Hybrid perturbation theory}

Using the Limber approximation and  neglecting the uncertainty in the cluster and galaxy redshift distributions, $\Sigma_{\rm g}(R)$ can be computed as:
\begin{equation}\label{eq:sigmag_PT}
    \Sigma_{\rm g}(R) = \bar{\Sigma}_{\rm g} \int dz \, \frac{n_{\rm g}(z) n_{\rm c}(z)}{d\chi/dz} \, w_p(R,z),
\end{equation}
where $n_{\rm c}(z)$ is the normalized redshift distribution of clusters  shown in Fig.~\ref{fig:cl-dist}, $n_{\rm g}(z)$ is the normalized redshift distribution of galaxies, which we obtain from the BPZ photo-z  estimates and 
\begin{equation}
w_p(R,z) = \int dl \, \xi_{\rm cg}\bigg( \sqrt{r^2_p + l^2} \bigg),
\end{equation}
where $\xi_{\rm cg}(r)$ is the 3D cluster galaxy correlation function. Using our hybrid PT-based model of galaxy bias,  $\xi_{\rm cg}$ can be written as:
\begin{eqnarray}\label{eq:xicg}
\xi_{\rm cg}(r) =
\begin{cases}
\frac{1}{A_{\rm norm}} \, \rho(r) \,, \ \ \ r < r_{\rm patch} \, \\
\xi^{\rm PT}_{\rm cg}(r) \,, \ \ \ ~r \geq r_{\rm patch} \,  
\, 
\end{cases}
\end{eqnarray}
where $\rho(r)$ is defined in Eq.~\ref{eq:rhoDK} and $A_{\rm norm} = \bar{\Sigma}_{\rm g} \int dz \, \frac{n_{\rm g}(z) n_{\rm c}(z)}{d\chi/dz}$ is a normalization which ensures that on small scales, the projected galaxy density estimate from this model converges to Eq.~\ref{eq:sigmag_theory}. We set $r_{\rm patch} = 15 \,h^{-1} {\rm Mpc}$.  
We test two different PT models for galaxy and cluster bias to estimate  ($\xi^{\rm PT}_{\rm cg}$): the first is the linear bias model, and the second expands the galaxy and cluster bias using 1-loop PT. Thus: 
\begin{eqnarray*}\label{eq:xiPT}
\xi^{\rm PT}_{\rm cg}(r) =
\begin{cases}
b^{(1)}_{\rm c} b^{(1)}_{\rm g} \xi^{\rm NL}_{\rm mm} \,, \ \  \hspace{1.1cm} \ ~{\rm Hybrid \, Linear \, PT \,  Bias \, model.} \,  \\
f(b^{(1)}_{\rm c},b^{(2)}_{\rm c},b^{\rm (s)}_{\rm c},b^{\rm (3nl)}_{\rm c}, b^{(1)}_{\rm g},b^{(2)}_{\rm g},b^{\rm (s)}_{\rm g},b^{\rm (3nl)}_{\rm g}, P^{\rm NL}_{\rm mm}, P^{\rm Lin}_{\rm mm}) \\  \hspace{3cm} {\rm Hybrid \, 1-loop \, PT \, Bias \, model} \,
\end{cases}
\end{eqnarray*}
The various bias parameters introduced above can be motivated as follows. 
Assuming isotropy and homogeneity, the overdensity of  biased tracers of  matter, such as clusters and galaxies, can be described using scalar quantities constructed from the matter density, the divergences of the  velocity and the gravitational potential field. At linear order, only the gravitational evolution of the matter density contributes which results in a simple expression for the cluster-galaxy correlation that depends only on their large scale linear biases, $b^{(1)}$. Including all the terms contributing up to third order results in the 1-loop PT prediction which adds three extra parameters: the second order tracer-bias $b^{\rm (2)}$, shear bias $b^{\rm (s)}$ and non-local bias $b^{\rm (3nl)}$ for both clusters and galaxies. The explicit functional form of the 1-loop PT prediction as implemented here follows  \citet{McDonald2009}, \citet{Saito2014a}, and \citet{Pandey_2020}.

The matter correlation function ($\xi^{\rm NL}_{\rm mm}$) used in our PT model is estimated using the \textsc{halofit} fitting function \citep{Takahashi:2012:}. 
We incorporate the effects of miscentering by using the same methodology as detailed in Sec.~\ref{sec:model}. 
We assume $b^{(1)}_{\rm c}=5.5$ and $b^{(1)}_{\rm g}=1.8$ as our fiducial choice of the effective linear bias of clusters and galaxies respectively. These bias values can in principle be obtained from the auto correlation of the clusters and galaxies, but here we obtain approximate values as follows. 
Our estimate of $b_{\rm c}$ is obtained from the  \cite{Tinker2010} fitting function for the mean mass and redshift of our cluster sample.
The galaxy bias is bounded from above by measurements of the auto-correlation of \texttt{redMaGiC} galaxies in  DES \citep{Rozo:2016:,Elvin-Poole:2018:} over the same redshift range. 
For the 1-loop PT curve, we assume the bias parameters $b^{\rm (s)}$ and $b^{\rm (3nl)}$ to be equal to their co-evolution value \citep{McDonald2009, Saito2014a} for both clusters and galaxies 
\footnote{The co-evolution value referred to here is obtained by equating the Eulerian and Lagrangian prescription of galaxy biasing, which results in $b^{\rm (s)} = \frac{-4}{7}(b^{(1)}-1)$ and $b^{\rm 3nl}=(b^{(1)}-1)$}. 
The second order bias parameter for clusters ($b^{\rm (2)}_{\rm c}$) is estimated using the relation calibrated from N-body simulations as described in \citet{Lazeyras_2016}. For the galaxies, we approximate the  second order bias using the calibration described in \citet{Pandey_2020} which used DES mocks.

The resulting model predictions are shown in the left panel of Fig.~\ref{fig:galaxy-profile}.
We show our fiducial result from Sec.~\ref{sec:result-gal} (blue), overplotted with the fit from the PT models.  
It is evident that both the PT models fit the data above 15$h^{-1} {\rm Mpc}$ and match the DK14 fit with the choice of bias parameters shown. Also note that the linear and 1-loop PT models differ at approximately the 10\% level; these differences are degenerate with the uncertainties in the photometric redshifts of galaxies and their bias values. 
We defer a more detailed modeling and validation of cluster-galaxy and cluster-matter correlations over all scales to a future study. 

\subsubsection{HOD-based models}

Our second approach is an analytical and simulation-based calculations of $\Sigma_{\rm g}$ using the halo occupation distribution (HOD) framework \citep[e.g.,][]{CooraySheth2002}. The results are shown in the right panel of Fig. \ref{fig:galaxy-profile}. As in \citet{Salcedo_et_al_2020} we recast the standard 5 parameter HOD in terms of the galaxy number density $n_\mathrm{gal}$ and ratios of the standard mass-parameters $(M_1 / M_\mathrm{min})$ and $(M_0 / M_1)$. We use the standard approach of modeling central and satellite galaxies' HODs separately \citep[e.g.,][]{Kravtsov_et_al_2004, Zheng_et_al_2005, Zehavi_et_al_2011}, which allows us to additionally model the incompleteness of the central galaxy sample: 
\begin{equation}
\langle N_{\rm g} | M_h \rangle = f_\mathrm{cen} \langle N_\mathrm{cen} | M_h \rangle + \langle N_\mathrm{sat} | M_h \rangle,
\end{equation}
where the parameter $f_\mathrm{cen}$ expresses the level of central incompleteness. We also allow the satellite galaxy profile to differ from that of the matter,
\begin{equation}
c_\mathrm{sp}^\mathrm{gal} = \mathrm{A}_\mathrm{con} \, c_\mathrm{sp}^\mathrm{halo}
\end{equation}
where $c_\mathrm{sp}^\mathrm{halo} = r_h / r_s$ is the halo concentration. Finally we characterize the cluster mass-observable relation as a linear relation with a constant lognormal scatter $\sigma_{\ln M_{\rm c}}$,
\begin{equation}
    \ln M_\mathrm{obs} = \ln M_h + \sigma_{\ln M_{\rm c}} \times \mathcal{N}(0,1).
\end{equation}
To analytically compute $\Sigma_{\rm g}$ we first compute the real-space cluster-galaxy correlation function $\xi_{\rm cg}$ as an effective sum of 1- and 2-halo terms,
\begin{equation}\label{eq:1h2hxi}
    \xi_{\rm cg} = ((\xi_{\rm cg}^\mathrm{1h})^{\alpha} + (\xi_{\rm cg}^\mathrm{2h})^{\alpha})^{1/\alpha},
\end{equation}
where $\alpha$ controls the smoothness of the transition between 1-halo and 2-halo contributions. We choose $\alpha = 2$ as our fiducial choice but also show the impact of changing this value; the choice $\alpha=2$ is similar to simply choosing the maximum of the two terms at a given radius per \citet{Hayashi08}. The line of sight integral of this total $\xi_{\rm cg}$ is used to obtain the predicted  $\Sigma_{\rm g}$. 
The 2-halo term is given by,
\begin{equation}
    \xi_{\rm cg}^\mathrm{2h} = b_{\rm g} b_{\rm c} \xi_{mm},
\end{equation}
where the respective bias factors are written as,
\begin{equation}
b_{\rm g} = \frac{1}{n_{\rm g}} \int_0^\infty d M_h \frac{d n_h}{d M_h} \langle N_{\rm g} | M_h \rangle b_h(M_h),
\end{equation}
\begin{equation}
b_{\rm c} = \frac{1}{n_{\rm c}} \int_0^\infty d M_h \frac{d n_h}{d M_h}  b_h(M_h) \psi (M_h),
\end{equation}
and $\psi(M_h)$ is the cluster selection function written as,
\begin{equation}
\psi(M_h) = \int d M_\mathrm{obs} P(M_\mathrm{obs} | M_h) H(M_\mathrm{obs} - M_\mathrm{obs}^\mathrm{min} )
\end{equation}
where $H$ is the Heaviside step function. The 1-halo term is given by,
\begin{align}
1 + \xi_{\rm cg}^\mathrm{1h} (r) &= \frac{1}{2 \pi r^2 n_{\rm g} n_{\rm c}} \int_0^{\infty} d M_h \left< N_\mathrm{sat} (M_h) | N_\mathrm{cen} = 1 \right> \nonumber \\
&\times I' \left( \frac{r}{R_\mathrm{sp} (M_h)} , c_\mathrm{gal}^\mathrm{sp} (M_h) \right) \frac{1}{R_\mathrm{sp} (M_h)} \frac{d n}{d M_h}  \psi( M_h),
\end{align}
where $I^\prime$ is the normalized radial distribution of galaxies within the splashback radius of the haloes. The splashback radius is equal to  $1.1 \times r_{200m}$ (see Table~\ref{tab:rsp_comp}), where $r_{200m}$ is the radius at which average interior density of the halo equals 200 times the mean mass density of the universe. In the calculation of these two terms we utilize the halo mass function $d n / d M_h$ of \citet{Tinker2008}, the halo bias function $b(M_h)$ of \citet{Tinker2010}, and the fitting formula for the linear matter power spectrum of \citet{Eisenstein_Hu_1998}. 

We also compute a simulation based estimate of $\Sigma_{\rm g}$.
To do this we populate haloes identified in the MDPL2 simulation with the {\sc{rockstar}} halo finder \citep{Behroozi_2013} from the $z\approx0.5$ particle snapshot (see Sec.~\ref{sec:data-sim}). 
The relatively high resolution of MDPL2 is required to accurately model $\Sigma_{\rm g}$ for our galaxy sample. 
Mock galaxies and cluster catalogs are created as in \citet{Salcedo_et_al_2020}. 
Using these mock catalogs we directly compute $\xi_{cg}$ using $\sc{corrfunc}$ \citep{Sinha_2017} and integrate to obtain $\Sigma_{\rm g}$.

The results in Fig. \ref{fig:galaxy-profile} (right panel) show that the HOD predictions are in reasonable agreement with the measured profile. There are scale dependent features (relative to the lensing profile) in the transition region between the 1-halo and 2-halo terms ($\sim$2 $h^{-1}{\rm Mpc}$) -- these show some tension with the measurements. The two  analytical predictions shown by the dashed curve correspond to different ways of combining the 1- and 2-halo terms. They  use $\alpha=1,2$ in Eqn. \ref{eq:1h2hxi}. The $\alpha=2$ case is very close to another alternative in which the larger of the two terms is used. We also note that the sharp feature near the virial radius is due the truncation of the 1-halo term; other  plausible prescriptions would alter this feature. 

The figure also shows that there are some differences between simulations and analytical HOD based predictions. We  include the analytically computed Gaussian errors for $w_p$ on our simulation curve \citep{Salcedo_et_al_2020}. Since $\Sigma_{\rm g} = \rho_{\rm g} w_p$ for a redshift snapshot, these error bars are meant to be representative of the error in modeling $\Sigma_{\rm g}$ from a single limited volume simulation. They are not meant to represent the true sample variance for $\Sigma_{\rm g}$ since they do not include contributions from the galaxy space density and non-Gaussian contributions at small scales. 

The detailed inferences about the HOD-based model await future work. For all the theoretical model comparisons, we also plan to compare with measurements in narrower redshift bins, and with careful characterization of the photo-z estimates for better comparison with the redshift-dependent model predictions.

\begin{figure}
\centering
\includegraphics[width=0.99\linewidth]{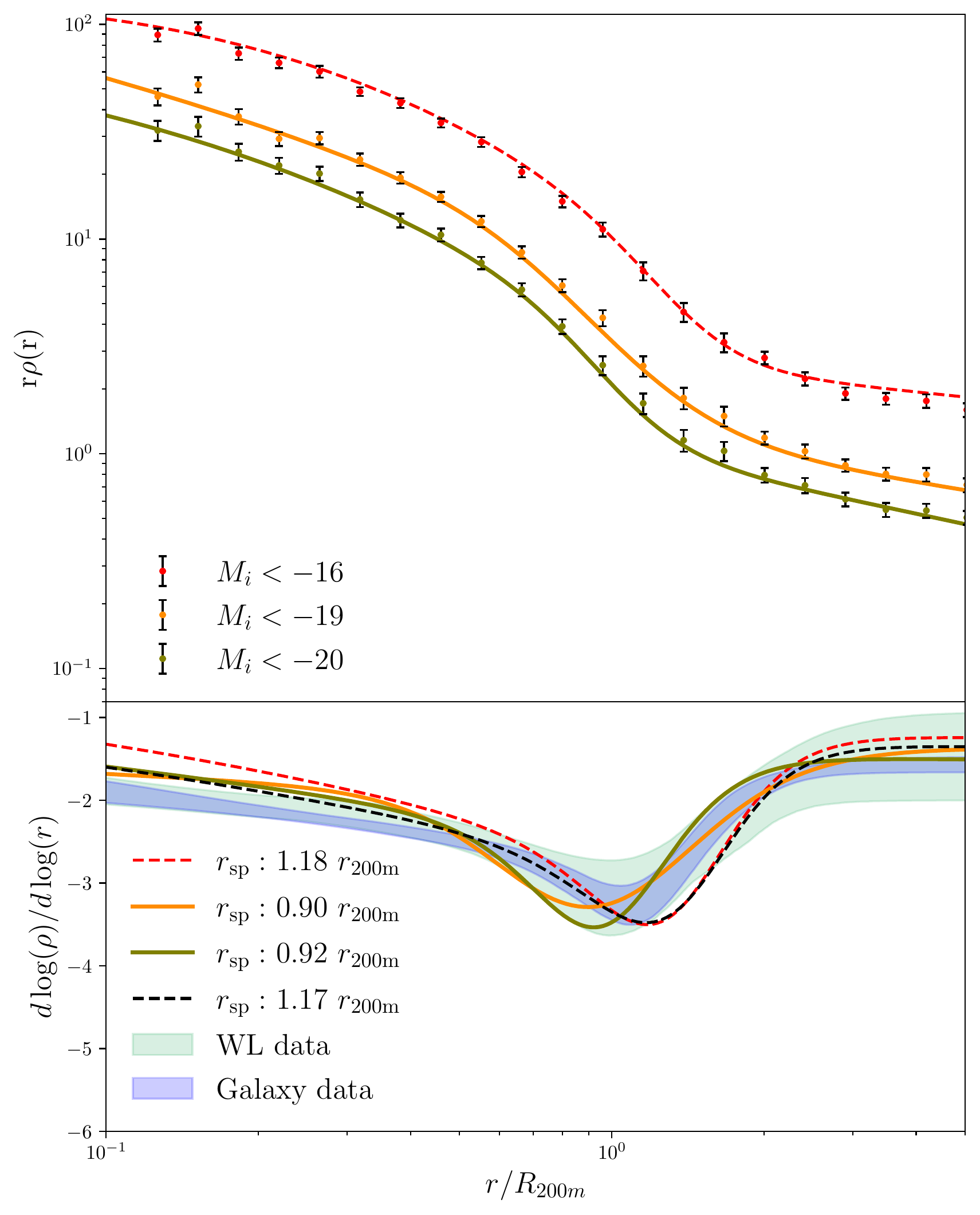}
\caption{
 Comparison of galaxy and mass profiles in the IllustrisTNG simulations. 
 \textit{Upper panel:} the 3D galaxy number density profiles around halos with mass ($M_{\rm 200m}>10^{14} h^{-1} M_{\odot}$)  for three different magnitude cuts (see legend), with the best-fit model fits shown as the smooth curves. The lower two samples (solid curves) bracket the selection of the galaxy sample from data. 
 \textit{Lower panel:} the logarithmic slope of the galaxy number density profiles from the upper panel. 
 Also shown are the slopes of the mass profiles (black dashed) from the same simulation, and the WL (green shaded region) and  galaxy (blue shaded region) profile fits from our data measurements.
 Note that the variable on the x-axis is $r/R_{\rm 200m}$ and the colors from the top panel transfer to the bottom panel.
 Although the clusters are somewhat less massive, the profiles are consistent with data for a similar galaxy luminosity cut.
 The splashback radii (see the legend of the bottom panel) are approximately 20\% smaller than that of the total matter in the IllustrisTNG simulations. For reference we also show the galaxy profile that includes essentially all the galaxies ($M_i<-16$): this profile is in closer agreement with the mass profile, as shown by \citet{TNG}. We discuss these further in Section \ref{sec:hydro_comp}. }
\label{fig:TNG}
\end{figure}

\subsection{Comparison with N-body and hydrodynamical  simulations}
\label{sec:hydro_comp}

Comparison of the predictions of cosmological simulations for the overall shape of both the radial matter density and galaxy number density profiles and our measurements in  Fig. \ref{fig:gal_wl_comp} shows that they agree well. Nevertheless, there are  hints of interesting  differences: the predicted dark matter profile slope, for example, is slightly shallower than the measured matter profile slope at $r\lesssim 1h^{-1}$ Mpc. The estimated  slope of the galaxy number density profile is also somewhat steeper than the predicted slope of the matter density profile at $r\lesssim 1h^{-1}$ Mpc.

The latter result is interesting because previous studies have generally concluded that the predicted number density profile of subhalos that are expected to host cluster satellite galaxies is less concentrated and {\it shallower} than the DM profile in the central regions. However, it was also recognized that the shape of the radial distribution of subhalos is sensitive to how subhalos are selected. For example, \citet{Nagai2004} used $N$-body$+$hydrodynamics simulations to show that the number density profile of galaxies is shallow if galaxies are selected by their subhalo mass, but is close to the DM profile if galaxies are selected by their stellar mass or by a subhalo property that is relatively insensitive to resolution and tidal stripping effects. 

This is consistent with more recent simulations as shown in Fig. \ref{fig:TNG}, where we compare our measurements with the profiles extracted from the IllustrisTNG simulations \citep{Pillepich18,Nelson19}. The mean cluster mass in this measurement is lower than in our study since the simulation does not have enough massive haloes to make a higher mass cut. The galaxy sample selection uses absolute magnitude $M_i<-19$ and $M_i<-20$ that bracket the magnitude cut of our sample.\footnote{See Appendix~\ref{app:galaxy_dimmer} for the comparison between the results for $M_i<-20$ and $M_i<-19$ galaxy samples.}
The shapes of the logarithmic slope profiles (bottom panel), in units of $r/r_{\rm 200m}$, are in reasonable agreement with our measurements. The shape of the galaxy profiles are also in reasonable agreement with the mass profile from the IllustrisTNG simulations. We note however, that the splashback radius measured using galaxies in the simulations is about 20\% smaller that that measured from the mass profile. This result is somewhat surprising as the subhalos that these galaxies inhabit do not show a smaller splashback in CDM-only simulations. By examining the phase space distribution of these massive galaxies, we find that the orbiting galaxies have higher density at smaller cluster-centric distance  compared to the dark matter and also have lower radial velocities. This can possibly be explained if the galaxies on radial orbits are preferentially tidally disrupted at pericenter and do not reach splashback. We defer a detailed study of this effect to future work. In a recent study of the splashback radius in IllustrisTNG \citep{TNG}, the galaxy sample used contained all objects that live in subhalos above $10^8 h^{-1} M_\odot$, including galaxies that are much fainter than our observed sample; for this sample they find that the splashback radius is in better agreement with that measured from the mass profile. We confirm this result in Fig.\ref{fig:TNG}, the red curve corresponding to the faintest simulation galaxy sample, has a splashback radius that agrees with the mass profile.

\citet{Budzynski2012} also showed that models based on cosmological $N$-body simulations that included modeling of ``orphan'' galaxies to account for the loss of subhalos in simulated catalogs due to resolution, also generally produced galaxy number density profiles close to those of matter or even steeper. Recently, \citet{Green21} presented a systematic study and calibration of the resolution and ``artificial disruption'' effects on the number density profile of subhalos in cosmological simulations. They showed that if resolution effects are corrected for, the subhalo number density profile is close to that of the matter distribution, while if a modest contribution of artificial disruption is additionally accounted for, the predicted number density profile is even mildly steeper than the matter density profile (see, e.g., their Fig. 7). 

Recent observational results of \citet{Shin19}, where we studied the galaxy distribution around SZ-selected clusters showed that the radial number density profile of galaxies  is at least as steep as the dark matter density profile from simulations. \citet{Shin19} has a similar cluster-mass selection as this work. The results presented here  are consistent with these results and are also consistent with recent simulation studies corrected for resolution effects. 

\subsection{Additional sources of  uncertainty}
\label{sec:systematics}

We discuss next several caveats regarding our analysis, which could be addressed in the future studies. 
First, the WL mass correction factors that are multiplied to the SZ-obtained mass \citep[see][]{Hilton_2020} are obtained from a comparison to richness-based mass calibration of DES Y3 \textsc{redMaPPer} clusters, using those matched between the ACT DR5 and the DES Y3. 
The correction factor is a constant independent of the S/N, and is calculated by taking the mean ratio of the richness-based WL mass and the SZ mass, using clusters with ${\rm S/N} > 6$. 
However, the true correction factor could be dependent on the S/N of the clusters. 
Also, when taking the mass-richness relation from the DES Y3, one introduces a large scatter between the richness and the mass. 

The use off jackknife resampling  to estimate our covariance matrix may also introduce additional uncertainty into our analysis.  It is known that the jackknife resampling method underestimates the covariance at large scales \citep{Norberg09,Singh17}. 
We have done tests by varying the number of jackknife patches to check that the effect is small for the range of scales used in our study.  Analytical methods could in principle be used to improve our covariance estimates at large scales.  

Besides, the uncertainty on the photometric redshift and systematics in the shear measurements such as blending of galaxies can induce multiplicative biases to the measured WL profile \citep[e.g.][]{McClintock19,Maccrann20}. 
Nevertheless, multiplicative biases could not alter the shape of the measured density profile.
Since the focus of this study is to constrain the shapes of the halo density profiles, our result is robust against such biases.

We have also explored the sensitivity of our results to the priors we impose on model parameters.  These priors are motivated from simulations \citep[e.g.][]{Diemer2014,More16}. In Appendix~\ref{app:prior_test} we test that relaxing these priors does not affect the main conclusions of our analysis.

To ensure we select a similar set of galaxies over the redshift range (0.15-0.7), we have applied an absolute magnitude cut ($M_i<-19.87$), using distance modulus.
However, since galaxy SEDs are not flat, two identical galaxies sharing the same SED but at different redshift will have different $i$-band magnitudes. 
Therefore, our absolute magnitude cut does not completely ensure similar selections of galaxies over redshift. 
This effect could be avoided via  accurate SED fitting and K-correction \citep{Kcorr}. 
However, given the  photo-$z$ uncertainties of the individual galaxies in our sample, such corrections are not applied. We leave further investigations of these effects to follow-up papers.

\section{Discussion}
\label{sec:discussion}

\subsection{Summary}

We have measured the weak lensing and projected galaxy number density profiles around SZ-selected clusters using data from ACT DR5 and DES Y3.
We compare the two profiles in Fig.~\ref{fig:gal_wl_comp} (Sec.~\ref{sec:measure-dsigma},~\ref{sec:measure-sigmag})
having fit them to the DK14  model \citep{Diemer2014} which in turn is based on N-body simulations. 
Our main findings are as follows.

We measure the splashback radius, $r_{\rm sp}$,  from the weak lensing mass profiles --  the first such measurement from SZ-selected clusters.  To reiterate, the splashback feature forms the boundary between the virialized and infalling matter around haloes. We find that the splashback radius measured from weak lensing is in good agreement with that measured from the galaxies in our sample of clusters, as shown in Fig.~\ref{fig:gal_wl_comp}. This figure also shows 
that the depth of the splashback feature agrees well between the two measurements. 

\begin{table*}
	\centering
	\begin{tabular}{c|ccccccc}
		Reference & Measurement & Sample & Mean mass[$10^{14}h^{-1}M_{\odot}$] & Mean redshift & $\frac{r_{\rm sp}^{\rm Optical}}{r_{\rm 200m}}$ & $\frac{r_{\rm sp}^{\rm X-ray}}{r_{\rm 200m}}$ & $\frac{r_{\rm sp}^{\rm SZ}}{r_{\rm 200m}}$ \\ \hline
		\citet{Baxter17} & galaxy profile & SDSS RM & $M_{\rm 200m}=1.9$ & 0.24 & $0.85\pm0.06$ & - & -   \\
        \citet{Chang2017} & galaxy profile & DES RM & $M_{\rm 200m}=1.8$ & 0.41 & $0.82\pm0.05$  & - & -\\ 
        \citet{Chang2017} & weak lensing &  DES RM & $M_{\rm 200m}=1.8$ & 0.41 & $0.97\pm0.15$  & - & -\\ \hline
        \citet{Murata20} & galaxy profile & HSC CAMIRA & $M_{\rm 200m} = 1.7$ & 0.57 & $1.14 \pm 0.14$ & - & - \\ \hline
        \citet{contigiani19} & weak lensing & CCCP X-ray & $M_{\rm 200m}=14$ & 0.28 & - & $1.34^{+0.45}_{-0.26}$ & -  \\
        \citet{Bianconi20} & galaxy profile & LoCuss X-ray & $M_{\rm 200m} = 14.1$ & 0.23 & - & $1.74 \pm 0.34$ & - \\ \hline
		\citet{Shin19} & galaxy profile & SPT SZ & $M_{\rm 200m} = 5.3$ & 0.49 & - & - & $1.22^{+0.26}_{-0.25}$ \\
        \citet{Shin19} & galaxy profile & ACT SZ & $M_{\rm 200m} = 5.8$ & 0.49 & - & - & $1.11^{+0.36}_{-0.28}$ \\
        \citet{Zuercher19} & galaxy profile & Planck SZ & $M_{\rm 200m} = 6.2$ & 0.18 & - & - & $0.92^{+0.13}_{-0.15}$ \\
        this work & weak lensing & ACT SZ & $M_{\rm 200m} = 4.8$ & 0.46 & - & - & $1.16^{+0.21}_{-0.29}$ \\
        this work & galaxy profile & ACT SZ & $M_{\rm 200m} = 4.8$ & 0.46 & - & - & $1.10^{+0.06}_{-0.14}$ \\ \hline
        this work & matter profile & MDPL2 $N$-body & $M_{\rm 200m} = 4.8$ & 0.48 & - & - & 1.07
	\end{tabular}
	\caption{The splashback radius $r_{\rm sp}$ from previous studies as well as this paper. Note that we normalize $r_{\rm sp}$ by $r_{\rm 200m}$ for easier comparison.
	}
	\label{tab:rsp_comp}
\end{table*}

The mass profile inferred from our lensing measurements is in agreement with the profiles of cluster haloes in $N$-body simulations. 
In contrast, measurements of $r_{\rm sp}$ around optically selected clusters  have been shown to be sensitive to parameters used in optical cluster selection methods \citep{More16,Chang2017,Baxter17, Murata20}. The profile shapes of these clusters have also shown possible evidence of selection effects as discussed in Sec. 1. In Table~\ref{tab:rsp_comp}, we summarize the measurements of the splashback radius from  previous studies.

In addition to the location of  splashback, the full profile of the projected mass agrees with that of galaxies over scales $\sim 0.2-20 h^{-1} {\rm Mpc}$. This result is somewhat surprising as nonlinear effects, such as merging and tidal disruption that can modify the radial distribution of galaxies are expected to alter the galaxy profile within and around the cluster halo, and scale dependent galaxy bias is expected to do so in the quasilinear regime at $\sim 10 h^{-1} {\rm Mpc}$. 

Our results may be understood in part by noting that clusters are rare peaks in the matter distribution that dominate their environment. The turn-around region around cluster mass haloes can extend up to $\sim 5 R_{vir}$, within which the galaxies and dark matter infall under the cluster potential. Hence their dynamics is determined primarily by the cluster's gravity, with the dark matter and galaxies acting as  tracers. Therefore, barring an overall scaling determined by the underlying bias of the galaxy sample relative to the dark matter, the shapes of the dark matter profile and the galaxy profile may be very similar, as observed. A detailed model for the intermediate infall region around haloes  based on the gravitational collapse of matter on to the halo is in preparation \citep{GarciaAdhikariRozoinprep}.

Within the splashback radius the agreement between the galaxy profile and dark matter profile has somewhat different implications. The bulk of the population (DM or galaxies) within splashback is orbiting in the cluster potential. Unlike dark matter, galaxies are expected to undergo tidal disruptions and merge into the central BCG. However, our results show that for cluster mass haloes a significant number of galaxies survive in the inner regions such that the overall population traces the dark matter on the scales considered in this paper. Note that the quenching of galaxies in the cluster environment changes their colors (see e.g. the difference in red and blue galaxy profiles in \cite{Adhikari2020}), but our results imply that the total number of galaxies in luminosity bins is largely preserved.  

We have shown preliminary results on  different approaches to modeling cluster-galaxy correlations (Sections~ \ref{sec:theory_comp} and \ref{sec:hydro_comp}). The three models we consider are: (1) DK14 combined with perturbation theory on large scales, (2) Standard HOD based halo modeling, and (3) Hydrodynamical simulations. The results are shown in Figs.~\ref{fig:galaxy-profile} and \ref{fig:TNG}. They demonstrate that the first of the three approaches can fit the measured profiles from deep inside the halo to the linear regime. The  scale dependence in the quasilinear regime  predicted by second order perturbation theory is testable with higher precision measurements. 
The second approach, based on the standard halo model and assigning galaxies using HODs, typically shows scale dependent features in the galaxy profile (relative to matter) in the transition from the 1- to 2-halo regime. However within our uncertainties, the HOD approach provides a reasonable fit to the data. The galaxy profiles from the IllustrisTNG hydrodynamical simulations (matched to our magnitude cuts) are also broadly consistent with the data, though with some modest deviations (Fig. \ref{fig:TNG}). We leave for future work a detailed investigation of these models. 

To summarize, we find that although galaxy clusters are highly nonlinear objects, their dominance of the gravitational field over significantly large distances may be responsible for the observed simplicity of galaxy and mass profiles.

\subsection{Future directions and caveats}

SZ-selected clusters offer promise in comparing measurements with theoretical predictions based on halo mass selection. Our results show the agreement between data and CDM simulations (along with the IllustrisTNG hydrodynamical simulations). The location of  splashback in  the galaxy and matter profiles agrees with that  from the  $N$-body simulations, with an uncertainty of about 20\% for WL and 9\% for galaxies. This must be further validated with detailed simulated analyses that represent the noise properties of the data, which can also test whether the true profiles of clusters are recovered independent of the cosmological model. 

The splashback radius can be used to define cluster masses. Since this radius represents the dynamical boundary of the cluster, it cleanly separates the true ``1--halo'' term. Moreover, \citet{Diemer2020} shows that defining the cluster boundary with this radius has the advantage of a more universal mass function.  Therefore, given a cluster sample binned in an observable such as SZ signal-to-noise, one can use the splashback radius to carry out its lensing mass calibration. Higher precision measurements from the full DES survey and other upcoming surveys will enable these applications, with improvements expected in cosmological analyses that use cluster counts and cluster clustering.

We have found that the mass and galaxy profiles of SZ-selected clusters are remarkably similar up to an overall scaling.
Further tests of this similarity using higher precision measurements are warranted.
If agreement between the two continues to hold,  the galaxy profile (which can be measured at higher signal-to-noise) can provide priors on the shape of the mass profile.  This could be useful for lensing mass calibration, because once the shape of the DM profile is set by the priors,  cluster mass estimation boils down to constraining only the amplitude of the DM profile.  Such priors could be particularly useful for obtaining halo masses from observations of lensing of the cosmic microwave background \citep[CMB;][]{Baxter:2015,Madhavacheril:2015}.  It is difficult to use low resolution CMB observations to constrain the shape of cluster mass profiles; a prior on the shape from galaxy observations would therefore likely improve CMB lensing mass constraints.   Surveys that do not have lensing measurements available may be able to use the galaxy profiles to obtain mass estimates of cluster samples, provided the galaxy population is similar to the ones studied here. 

\citet{Diemer2014} and subsequent work also show that the location of the splashback radius normalized by $r_{\rm 200m}$ is a strong function of the mass accretion rate of the clusters. Given the measurements of the splashback radius over a wide redshift range, we can extend our parameter space to incorporate a model for the accretion rate dependence of the splashback radius and constrain the distribution of the mass accretion rates of different cluster samples; this can help us better understand the halo formation history and construct accurate halo models for cluster studies.

Various astrophysical processes inside clusters such as tidal disruption, dynamical friction and star formation quenching can alter the shape of the galaxy profiles. 
By splitting the galaxy and the cluster sample into  bins of, for example, galaxy magnitude, galaxy color, cluster redshift, cluster mass and cluster environment, we can test how these processes affect the formation of galaxy profiles and their relationship to the DM halo as well as to the large scale structure. 
By comparing such  measurements with $N$-body/hydrodynamical simulations and models, one can better understand the physics involved in halo formation. 

The mass profiles measured  using weak lensing can be used to constrain baryonic models on  cluster scales \citep{Schneider19, 2019MNRAS.488.1652H}. The mass  profiles are also sensitive to dark matter models such as self-interacting dark matter (SIDM) models. SIDM can increase the central density of the dark matter profile within the scale radius and cause  it to steepen significantly in the region between the scale radius and the virial radius of the halo \citep{Banerjee20}. While our mass profiles are currently statistically consistent with cold dark matter, we note that a steeper profile between $0.2$ and $1$ Mpc $h^{-1}$ as expected for realistic SIDM models is not ruled out by our current measurements. We defer a detailed investigation of the constraints obtained from current measurements  to a future study.

Our preliminary results on models for cluster-galaxy correlations are encouraging, in that the models describe the galaxy and mass profiles reasonably well. This motivates detailed theoretical investigations to produce accurate models that work for different galaxy samples and clusters across some range in mass and redshift. The redshift evolution is not an issue we have investigated beyond the redshift split test shown in the Appendix~\ref{app:split}; this can be done better with the full DES survey data. 

Sec. 4.6 discusses additional sources of uncertainty in our measurement and models. Beyond these, open questions remain about the universality of the galaxy profile shape for samples with different luminosity and redshift cuts. The DK14 model has a large number of free parameters; exploring tighter priors based on theory is a useful exercise for future work. Finally, the agreement with lensing is only as good as the lensing error bars, which exceed 20\% beyond about 10 Mpc. Improved measurements will enable more stringent tests at both large and small scales.

\section*{Acknowledgements}

We thank Neal Dalal and Ravi Sheth for stimulating discussions. 
We are grateful to Phil Mansfield, Chun-Hao To for comments on the draft.

The CosmoSim database used in this paper is a service by the Leibniz-Institute for Astrophysics Potsdam (AIP).
The MultiDark database was developed in cooperation with the Spanish MultiDark Consolider Project CSD2009-00064.

We gratefully acknowledge the Gauss Centre for Supercomputing e.V. (www.gauss-centre.eu) and the Partnership for Advanced Supercomputing in Europe (PRACE, www.prace-ri.eu) for funding the MultiDark simulation project by providing computing time on the GCS Supercomputer SuperMUC at Leibniz Supercomputing Centre (LRZ, www.lrz.de).

Funding for the DES Projects has been provided by the U.S. Department of Energy, the U.S. National Science Foundation, the Ministry of Science and Education of Spain, 
the Science and Technology Facilities Council of the United Kingdom, the Higher Education Funding Council for England, the National Center for Supercomputing 
Applications at the University of Illinois at Urbana-Champaign, the Kavli Institute of Cosmological Physics at the University of Chicago, 
the Center for Cosmology and Astro-Particle Physics at the Ohio State University,
the Mitchell Institute for Fundamental Physics and Astronomy at Texas A\&M University, Financiadora de Estudos e Projetos, 
Funda{\c c}{\~a}o Carlos Chagas Filho de Amparo {\`a} Pesquisa do Estado do Rio de Janeiro, Conselho Nacional de Desenvolvimento Cient{\'i}fico e Tecnol{\'o}gico and 
the Minist{\'e}rio da Ci{\^e}ncia, Tecnologia e Inova{\c c}{\~a}o, the Deutsche Forschungsgemeinschaft and the Collaborating Institutions in the Dark Energy Survey. 

The Collaborating Institutions are Argonne National Laboratory, the University of California at Santa Cruz, the University of Cambridge, Centro de Investigaciones Energ{\'e}ticas, 
Medioambientales y Tecnol{\'o}gicas-Madrid, the University of Chicago, University College London, the DES-Brazil Consortium, the University of Edinburgh, 
the Eidgen{\"o}ssische Technische Hochschule (ETH) Z{\"u}rich, 
Fermi National Accelerator Laboratory, the University of Illinois at Urbana-Champaign, the Institut de Ci{\`e}ncies de l'Espai (IEEC/CSIC), 
the Institut de F{\'i}sica d'Altes Energies, Lawrence Berkeley National Laboratory, the Ludwig-Maximilians Universit{\"a}t M{\"u}nchen and the associated Excellence Cluster Universe, 
the University of Michigan, NFS's NOIRLab, the University of Nottingham, The Ohio State University, the University of Pennsylvania, the University of Portsmouth, 
SLAC National Accelerator Laboratory, Stanford University, the University of Sussex, Texas A\&M University, and the OzDES Membership Consortium.

Based in part on observations at Cerro Tololo Inter-American Observatory at NSF's NOIRLab (NOIRLab Prop. ID 2012B-0001; PI: J. Frieman), which is managed by the Association of Universities for Research in Astronomy (AURA) under a cooperative agreement with the National Science Foundation.

The DES data management system is supported by the National Science Foundation under Grant Numbers AST-1138766 and AST-1536171.
The DES participants from Spanish institutions are partially supported by MICINN under grants ESP2017-89838, PGC2018-094773, PGC2018-102021, SEV-2016-0588, SEV-2016-0597, and MDM-2015-0509, some of which include ERDF funds from the European Union. IFAE is partially funded by the CERCA program of the Generalitat de Catalunya.
Research leading to these results has received funding from the European Research
Council under the European Union's Seventh Framework Program (FP7/2007-2013) including ERC grant agreements 240672, 291329, and 306478.
We  acknowledge support from the Brazilian Instituto Nacional de Ci\^encia
e Tecnologia (INCT) do e-Universo (CNPq grant 465376/2014-2).

This manuscript has been authored by Fermi Research Alliance, LLC under Contract No. DE-AC02-07CH11359 with the U.S. Department of Energy, Office of Science, Office of High Energy Physics.

The ACT project is supported by the U.S. National Science Foundation through awards AST-1440226, AST-0965625 and AST-0408698, as well as awards PHY-1214379 and PHY-0855887. Funding was also provided by Princeton University, the University of Pennsylvania, and a Canada Foundation for Innovation (CFI) award to UBC. ACT operates in the Parque Astron\'{o}mico Atacama in northern Chile under the auspices of the Comisi\'{o}n Nacional de Investigaci\'{o}n Cient\'{i}fica y Tecnol\'{o}gica de Chile (CONICYT). Computations were performed on the GPC supercomputer at the SciNet HPC Consortium and on the hippo cluster at the University of KwaZulu-Natal. SciNet is funded by the CFI under the auspices of Compute Canada, the Government of Ontario, the Ontario Research Fund - Research Excellence; and the University of Toronto. The development of multichroic detectors and lenses was supported by NASA grants NNX13AE56G and NNX14AB58G.

Work at Argonne National Laboratory was supported under U.S. Department of Energy contract DEAC02-06CH11357.

JPH acknowledges funding for SZ cluster studies from NSF AAG number
AST-1615657.

Research at Perimeter Institute is supported in part by the Government of Canada through the Department of Innovation, Science and Industry Canada and by the Province of Ontario through the Ministry of Colleges and Universities.

KM acknowledges support from the National Research Foundation of South Africa.

\label{lastpage}

\bibliographystyle{mn2e_adsurl}
\bibliography{splash}

\section*{Affiliations}
$^{1}$ Department of Physics and Astronomy, University of Pennsylvania, Philadelphia, PA 19104, USA\\
$^{2}$ Kavli Institute for Cosmological Physics, University of Chicago, Chicago, IL 60637, USA\\
$^{3}$ Department of Astronomy and Astrophysics, University of Chicago, Chicago, IL 60637, USA\\
$^{4}$ Institute for Astronomy, University of Hawaii, Honolulu, HI 96822, USA \\
$^{5}$ Department of Astronomy and Center for Cosmology and
AstroParticle Physics, The Ohio State University, Columbus,
OH 43210, USA \\
$^{6}$ Department of Astronomy, Cornell University, Ithaca, NY 14853, USA \\
$^{7}$ Enrico Fermi Institute, University of Chicago, Chicago, IL 60637, USA \\
$^{8}$ Max Planck Institute for Extraterrestrial Physics, Giessenbach-strasse, 85748 Garching, Germany \\
$^{9}$ Faculty of Physics, Ludwig-Maximilians-Universit\"at, Scheinerstr. 1, 81679 Munich, Germany \\
$^{10}$ Cerro Tololo Inter-American Observatory, NSF's National Optical-Infrared Astronomy Research Laboratory, Casilla 603, La Serena, Chile \\
$^{11}$ Departamento de F\'isica Matem\'atica, Instituto de F\'isica, Universidade de S\~ao Paulo, CP 66318, S\~ao Paulo, SP, 05314-970, Brazil \\
$^{12}$ Laborat\'orio Interinstitucional de e-Astronomia - LIneA, Rua Gal. Jos\'e Cristino 77, Rio de Janeiro, RJ - 20921-400, Brazil \\
$^{13}$ Argonne National Laboratory, Lemont, IL 60439, USA \\
$^{14}$ Fermi National Accelerator Laboratory, P. O. Box 500, Batavia, IL 60510, USA \\
$^{15}$ Kavli Institute for Particle Astrophysics \& Cosmology, P. O. Box 2450, Stanford University, Stanford, CA 94305, USA \\
$^{16}$ Instituto de F\'{i}sica Te\'orica, Universidade Estadual Paulista, S\~ao Paulo, Brazil \\
$^{17}$ Institute of Cosmology and Gravitation, University of Portsmouth, Portsmouth, PO1 3FX, UK \\
$^{18}$ Physics Department, University of Wisconsin-Madison, Madison, WI  53706-1390, USA \\
$^{19}$ CNRS, UMR 7095, Institut d'Astrophysique de Paris, F-75014, Paris, France \\
$^{20}$ Sorbonne Universit\'es, UPMC Univ Paris 06, UMR 7095, Institut d'Astrophysique de Paris, F-75014, Paris, France \\
$^{21}$ Canadian Institute for Theoretical Astrophysics, University of Toronto, Toronto, ON M5S 3H4, Canada \\
$^{22}$ Department of Physics \& Astronomy, University College London, Gower Street, London, WC1E 6BT, UK \\
$^{23}$ Department of Physics, Carnegie Mellon University, Pittsburgh, Pennsylvania 15312, USA \\
$^{24}$ Department of Physics, Duke University Durham, NC 27708, USA \\
$^{25}$ Instituto de Astrofisica de Canarias, E-38205 La Laguna, Tenerife, Spain\\
$^{26}$ Universidad de La Laguna, Dpto. Astrofísica, E-38206 La Laguna, Tenerife, Spain \\
$^{27}$ Center for Astrophysical Surveys, National Center for Supercomputing Applications, Urbana, IL 61801, USA \\
$^{28}$ Department of Astronomy, University of Illinois at Urbana-Champaign, Urbana, IL 61801, USA \\
$^{29}$ Institut de F\'{\i}sica d'Altes Energies (IFAE), The Barcelona Institute of Science and Technology, Campus UAB, 08193 Bellaterra (Barcelona) Spain \\
$^{30}$ Lawrence Berkeley National Laboratory, Berkeley, CA 94720, USA \\
$^{31}$ Astronomy Unit, Department of Physics, University of Trieste, via Tiepolo 11, I-34131 Trieste, Italy \\
$^{32}$ INAF-Osservatorio Astronomico di Trieste, via G. B. Tiepolo 11, I-34143 Trieste, Italy \\
$^{33}$ Institute for Fundamental Physics of the Universe, Via Beirut 2, 34014 Trieste, Italy \\
$^{34}$ Observat\'orio Nacional, Rua Gal. Jos\'e Cristino 77, Rio de Janeiro, RJ - 20921-400, Brazil \\
$^{35}$ Department of Physics, The Ohio State University, Columbus, OH 43210, USA \\
$^{36}$ Department of Physics, IIT Hyderabad, Kandi, Telangana 502285, India \\
$^{37}$ Centro de Investigaciones Energ\'eticas, Medioambientales y Tecnol\'ogicas (CIEMAT), Madrid, Spain \\
$^{38}$ Physics Department, Stanford University, Stanford, CA 94305, USA \\
$^{39}$ Santa Cruz Institute for Particle Physics, Santa Cruz, CA 95064, USA \\
$^{40}$ Institute of Theoretical Astrophysics, University of Oslo. P.O. Box 1029 Blindern, NO-0315 Oslo, Norway \\
$^{41}$ Jet Propulsion Laboratory, California Institute of Technology, Pasadena, CA 91109, USA \\
$^{42}$ Department of Physics, Cornell University, Ithaca, NY 14853, USA \\
$^{43}$ Institut d'Estudis Espacials de Catalunya (IEEC), 08034 Barcelona, Spain \\
$^{44}$ Institute of Space Sciences (ICE, CSIC),  Campus UAB, Carrer de Can Magrans, s/n,  08193 Barcelona, Spain \\
$^{45}$ Department of Astronomy, University of Michigan, Ann Arbor, MI 48109, USA \\
$^{46}$ Department of Physics, University of Michigan, Ann Arbor, MI 48109, USA \\
$^{47}$ SLAC National Accelerator Laboratory, Menlo Park, CA 94025, USA \\
$^{48}$ Department of Physics, University of Oxford, Denys Wilkinson Building, Keble Road, Oxford OX1 3RH, UK \\
$^{49}$ Jodrell Bank Center for Astrophysics, School of Physics and Astronomy, University of Manchester, Oxford Road, Manchester, M13 9PL, UK \\
$^{50}$ Department of Astronomy, University of Geneva, ch. d'\'Ecogia 16, CH-1290 Versoix, Switzerland \\
$^{51}$ Department of Physics, Columbia University, New York, NY 10027, USA \\
$^{52}$ Center for Computational Astrophysics, Flatiron Institute, New York, NY 10010, USA \\
$^{53}$ Astrophysics Research Centre, University of KwaZulu-Natal, Westville Campus, Durban 4041, South Africa \\
$^{54}$ School of Mathematics, Statistics \& Computer Science, University of KwaZulu-Natal, Westville Campus, Durban4041, South Africa \\
$^{55}$ School of Mathematics and Physics, University of Queensland,  Brisbane, QLD 4072, Australia \\
$^{56}$ Department of Physics and Astronomy, Rutgers, the State University of New Jersey, Piscataway, NJ 08854-8019, USA \\
$^{57}$ Center for Astrophysics $\vert$ Harvard \& Smithsonian, Cambridge, MA 02138, USA \\
$^{58}$ Department of Physics, Yale University, New Haven, CT 06520, USA \\
$^{59}$ Department of Astronomy/Steward Observatory, University of Arizona, Tucson, AZ 85721-0065, USA \\
$^{60}$ Australian Astronomical Optics, Macquarie University, North Ryde, NSW 2113, Australia \\
$^{61}$ Lowell Observatory, Flagstaff, AZ 86001, USA \\
$^{62}$ David A. Dunlap Department of Astronomy \& Astrophysics, University of Toronto, Toronto, ON M5S 3H4, Canada \\
$^{63}$ Dunlap Institute of Astronomy \& Astrophysics, Toronto, ON M5S 3H4, Canada \\
$^{64}$ Department of Applied Mathematics and Theoretical Physics, University of Cambridge, Cambridge CB3 0WA, UK \\
$^{65}$ Perimeter Institute for Theoretical Physics, Waterloo ON N2L 2Y5, Canada \\
$^{66}$ Department of Physics, University of Chicago, Chicago, IL 60637, USA \\
$^{67}$ Department of Astrophysical Sciences, Princeton University, Peyton Hall, Princeton, NJ 08544, USA \\
$^{68}$ Instituci\'o Catalana de Recerca i Estudis Avan\c{c}ats, E-08010 Barcelona, Spain \\
$^{69}$ Department of Physics, University of Milano-Bicocca, 20126 Milano, Italy \\
$^{70}$ Instituto de F\'isica Gleb Wataghin, Universidade Estadual de Campinas, 13083-859, Campinas, SP, Brazil \\
$^{71}$ Kavli Institute at Cornell for Nanoscale Science, Cornell University, Ithaca, NY 14853, USA \\
$^{72}$ Joseph Henry Laboratories of Physics, Jadwin Hall, Princeton University, Princeton, NJ 08544, USA \\
$^{73}$ Department of Physics and Astronomy, Haverford College, Haverford, PA 19041, USA \\
$^{74}$ Institute of Astronomy, University of Cambridge, Madingley Road, Cambridge CB3 0HA, UK \\
$^{75}$ Department of Physics and Astronomy, Pevensey Building, University of Sussex, Brighton, BN1 9QH, UK
$^{76}$ Instituto de F\'\i sica, UFRGS, Caixa Postal 15051, Porto Alegre, RS - 91501-970, Brazil \\
$^{77}$ Department of Physics, California Institute of Technology, Pasadena, CA 91125, USA \\
$^{78}$ Brookhaven National Laboratory, Bldg 510, Upton, NY 11973, USA \\
$^{79}$ Instituto de F\'isica, Pontificia Universidad Cat\'olica de Valpara\'iso, Casilla 4059, Valpara\'iso, Chile \\
$^{80}$ School of Physics and Astronomy, University of Southampton,  Southampton, SO17 1BJ, UK \\
$^{81}$ Computer Science and Mathematics Division, Oak Ridge National Laboratory, Oak Ridge, TN 37831 \\
$^{82}$ NASA/Goddard Space Flight Center, Greenbelt, MD 20771, USA \\

\appendix

\section{Sample split tests of the galaxy density profiles}
\label{app:split}

\begin{figure}
\centering
\includegraphics[width=0.99\linewidth]{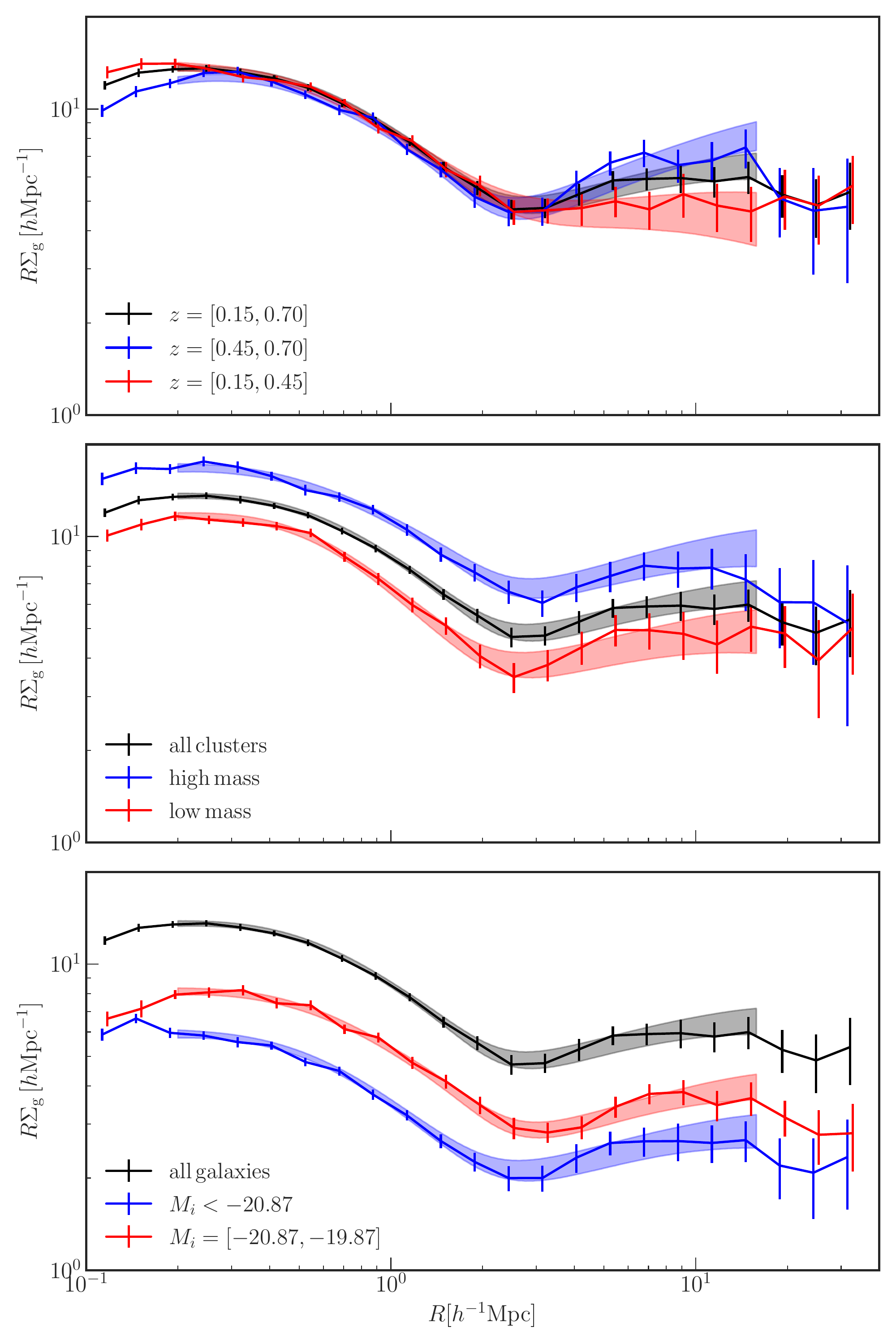}
\caption{The projected galaxy number density profiles around the ACT DR5 clusters. The shaded regions represent the 1-$\sigma$ ranges from the MCMC model fitting (Sec.~\ref{sec:model-fitting}).
In each panel, the black curve represents our fiducial SNR$>$4, z=[0,15,0.7]) measurement. \textit{top:} Projected galaxy number density profiles of the clusters split by redshift at 0.45: z=[0.15,0.45] (red) and z=[0.45,0.70] (blue). \textit{middle:} Projected galaxy number density profiles of the clusters split by mass: the high mass sample ($\langle M_{\rm 500c}\rangle = 3.75\times10^{14}h^{-1}M_{\odot}$, red) and the low mass sample ($\langle M_{\rm 500c}\rangle = 2.14\times10^{14}h^{-1}M_{\odot}$, blue). \textit{bottom:} Projected galaxy number density profiles of the clusters with galaxies split by magnitude: the low-luminosity sample ($-20.87<M_i<-19.87$, red) and the high-luminosity sample ($M_i<-20.87$, blue).
}
\label{fig:measurement-split}
\end{figure}

To test the robustness of the agreement between mass and light profiles, we split the galaxy sample by absolute magnitude, and the entire sample by redshift and cluster mass. The results are shown in Fig. \ref{fig:measurement-split}. 

We split our galaxies into a low luminosity sample ($M_i=[-20.87,-19.87]$) and a high luminosity sample ($M_i < -20.87$) and calculate the projected galaxy number density profiles  for each. The results are shown in the bottom panel of Fig.~\ref{fig:measurement-split}. We note that the shape of the radial profiles do not differ significantly between the two luminosity bins beyond $\sim 0.3 h^{-1} {\rm Mpc}$. At small radii the profile of the fainter sample becomes shallower compared to the bright sample.
Although we defer a detailed investigation of this effect to future work, we note here that these differences can arise from a few possible physical or systematic effects, including: low mass galaxies could be getting preferentially disrupted or merging into the BCG, and/or smaller/fainter galaxies are increasingly harder to detect at small radii due to the ambient light from the BCG. 
The location of the splashback radius agrees between the two samples.

We also split our cluster samples into a low (z=[0.15,0.45]) and  high (z=[0.45,0.7]) redshift sample. 
The results are shown in the top panels of Fig.~\ref{fig:measurement-split}.
There are  differences in the two profiles: the slope of the low redshift cluster sample tends to be steeper in the central and the infall regions and shallower at the location of the splashback.

Finally, we split our cluster samples into  high and  low mass samples, at the median mass. 
The results are shown in the middle panel of Fig.~\ref{fig:measurement-split}. There is no significant difference between these two samples. 
We leave further analysis of such tests to future studies. 

\section{The profile of dimmer galaxies}
\label{app:galaxy_dimmer}

\begin{figure}
\centering
\includegraphics[width=0.99\linewidth]{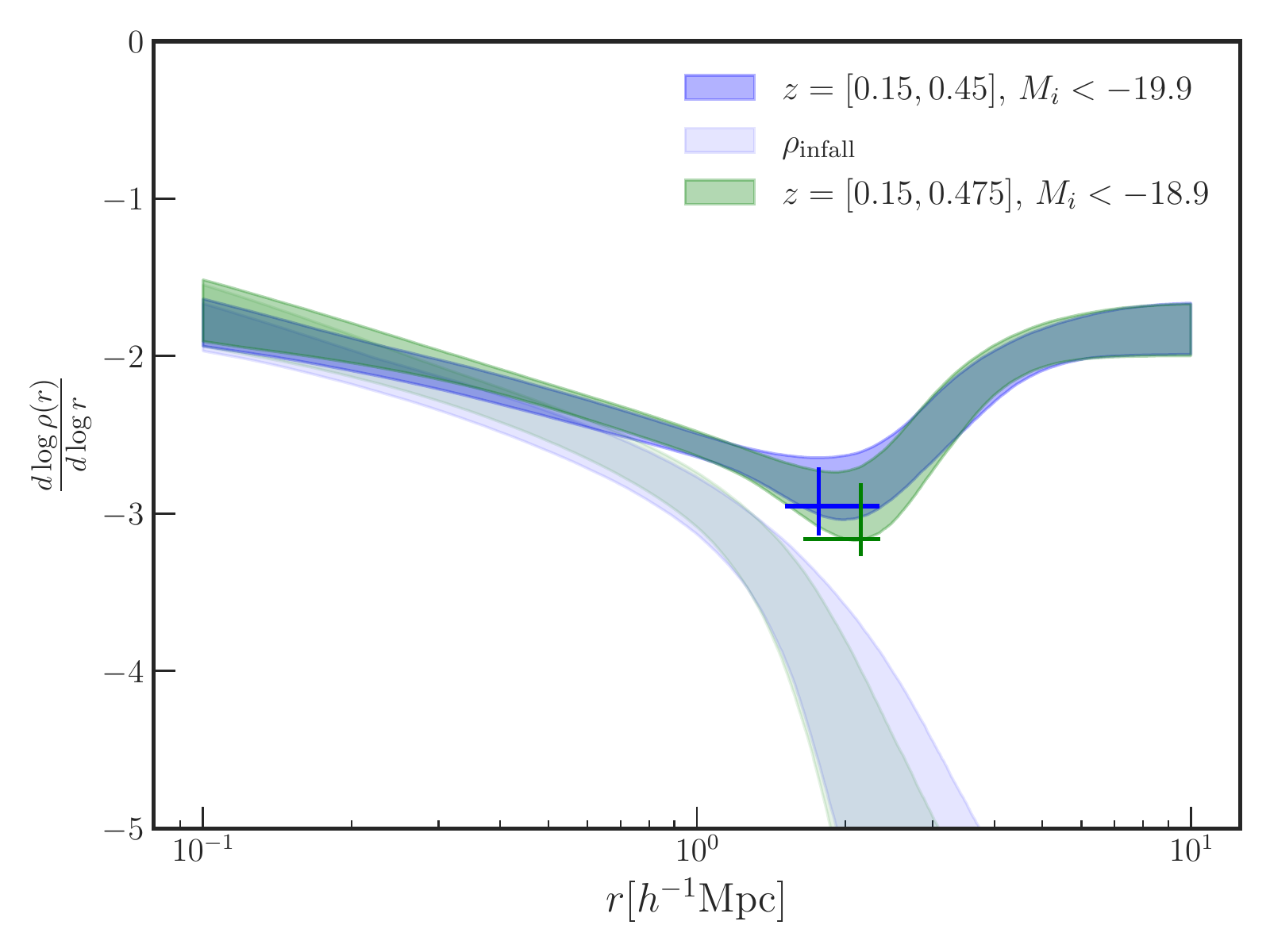}
\caption{The logarithmic slope of the profile of the low-z sample (see Sec.~\ref{app:split}) and the profile of galaxies with $M_i<-18.85$ (one magnitude dimmer than the fiducial sample) in the same redshift range.  
}
\label{fig:slope_dimmer}
\end{figure}

So far we have used galaxies with $M_i<-19.87$ as our fiducial choice. 
This absolute magnitude cut corresponds to the apparent magnitude cut ($m_i<22.5$) at the maximum redshift ($z=0.7$). 
If we limit our maximum redshift to $z=0.475$, $m_i<22.5$ corresponds to the absolute magnitude cut  $M_i<-18.85$, about one magnitude dimmer than our fiducial galaxies. 
It allows us to examine dimmer galaxies, and to make a comparison with the profiles from IllustrisTNG simulations shown in Fig. \ref{fig:TNG}. 
The result for the slope profile of this dim sample is shown in Fig.~\ref{fig:slope_dimmer} and compared to our fiducial sample using a similar maximum redshift ($z=0.45$). 
The two profiles exhibit similar shapes and splashback radii, unlike the case for IllustrisTNG simulations, in which the density profiles of galaxies with $M_i<-19$ and $M_i<-20$ display somewhat discrepant shapes.

\section{Effect of miscentering and other model parameters}
\label{app:prior_test}

\begin{figure}
\centering
\includegraphics[width=0.99\linewidth]{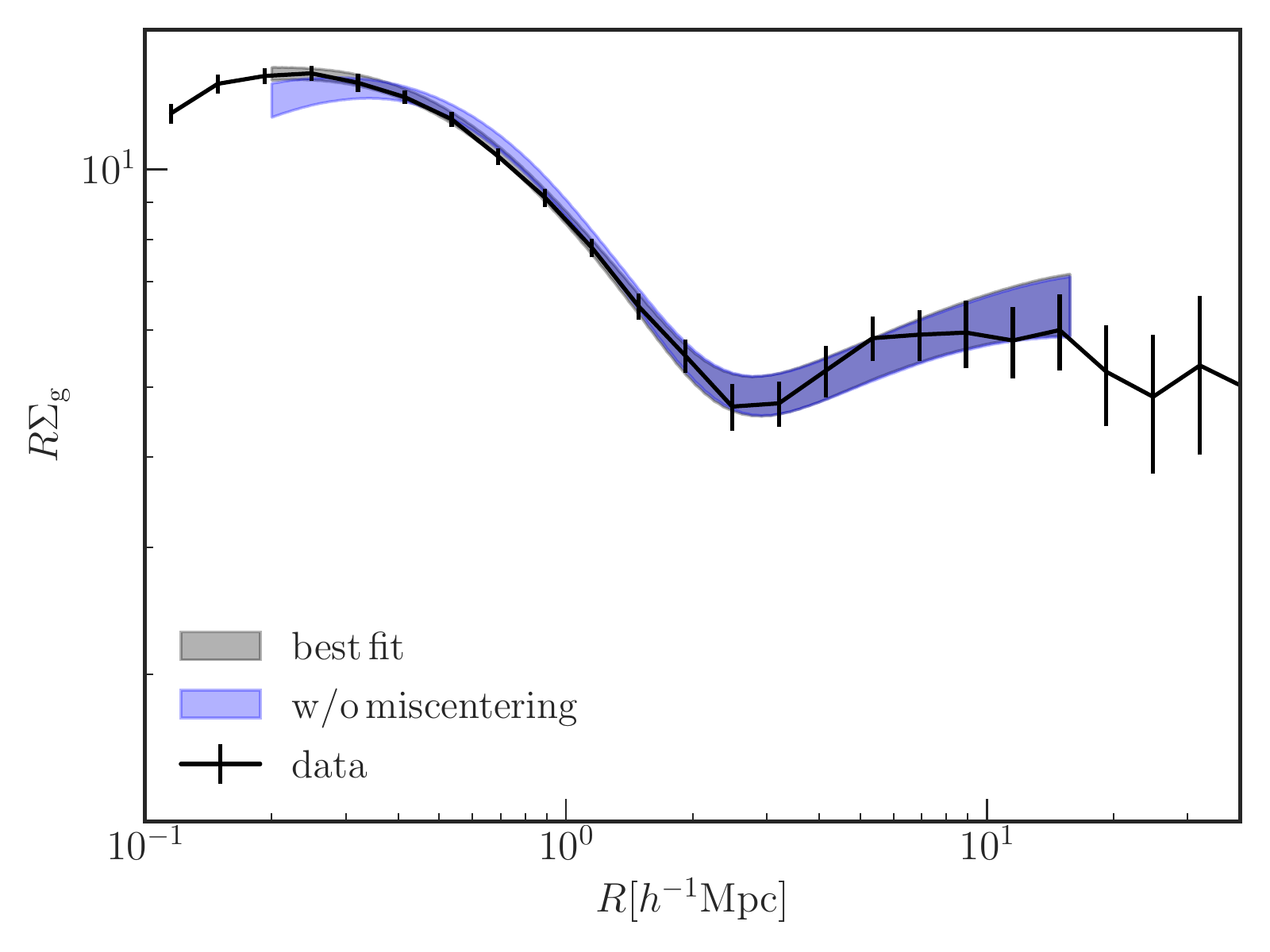}
\caption{The measurement of the projected galaxy number density profile (black, $\Sigma_{\rm g}$) and the 1-$\sigma$ range of the best fit from the MCMC chain (grey shade). The blue shade represents the 1-$\sigma$ range from the MCMC chain assuming no miscentering. One can confirm that  miscentering affects  the fitting only in the  central region ($R < 0.4 h^{-1} {\rm Mpc}$)
}
\label{fig:miscentering}
\end{figure}

The miscentering parameters for the galaxy and lensing profiles may differ somewhat --  the uncertainty on the lensing parameters is large so the differences in the parameters are not statistically significant. 
We have compared the measured profiles (with no miscentering correction) to the best fits and found that miscentering affects only the first 2-3 points (see figure \ref{fig:miscentering}). 
So, even if there was an error in the miscentering, the agreement of the inferred 3D density profiles would hold beyond the first few bins.   

\begin{figure}
\centering
\includegraphics[width=0.99\linewidth]{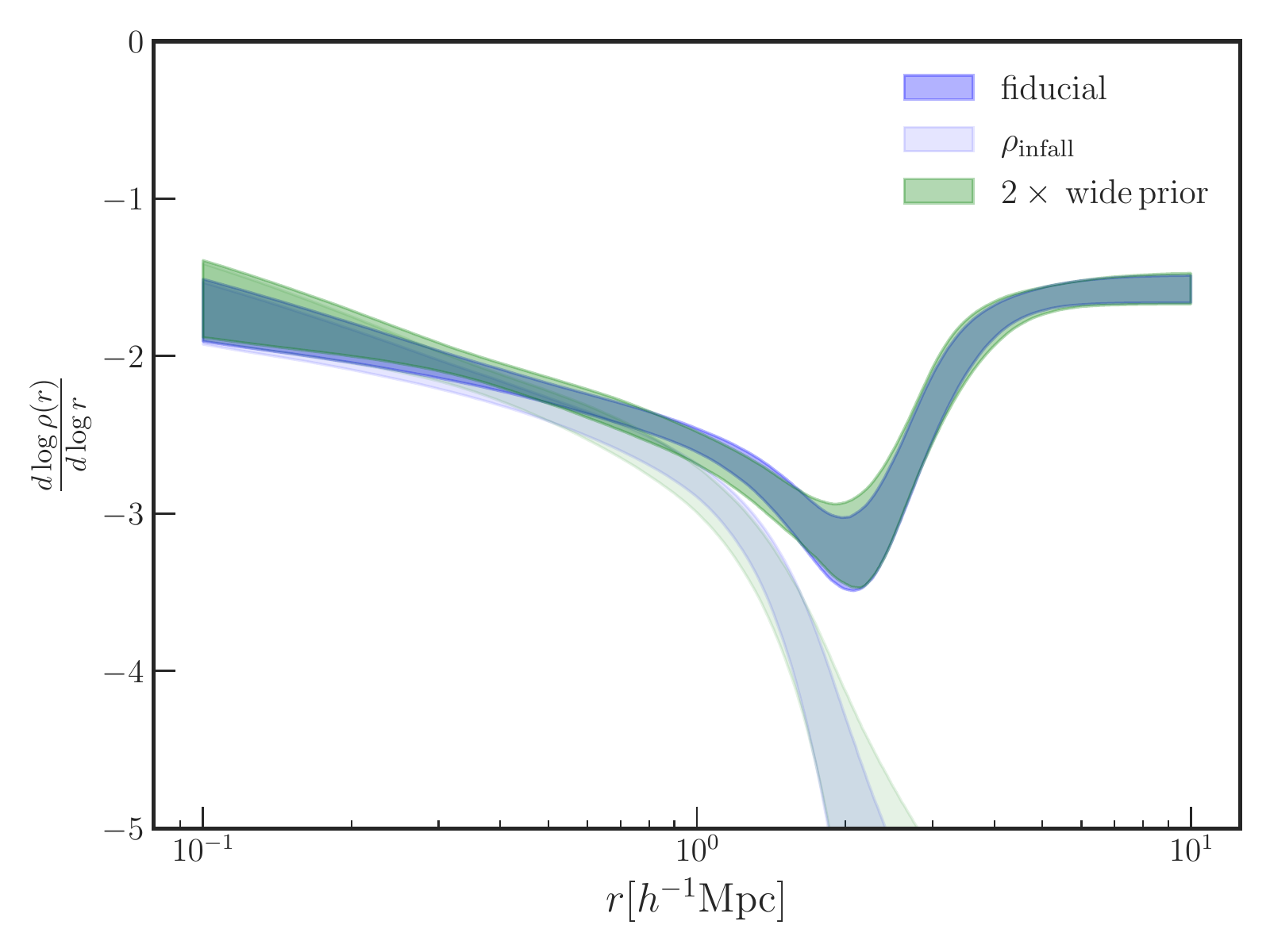}
\caption{The comparison of the fitted logarithmic slope profiles between the fiducial model (red) and the model with 2 times wider priors (green).
}
\label{fig:slope_prior}
\end{figure}

To test how much the priors (see Table~\ref{tab:modeling_parameters}) affect our fitting result, we apply 2 times wider priors to the MCMC chain. 
The result is shown in the Fig.~\ref{fig:slope_prior}. 
We confirm that the wider priors do not change our result significantly.

\begin{figure}
\centering
\includegraphics[width=0.99\linewidth]{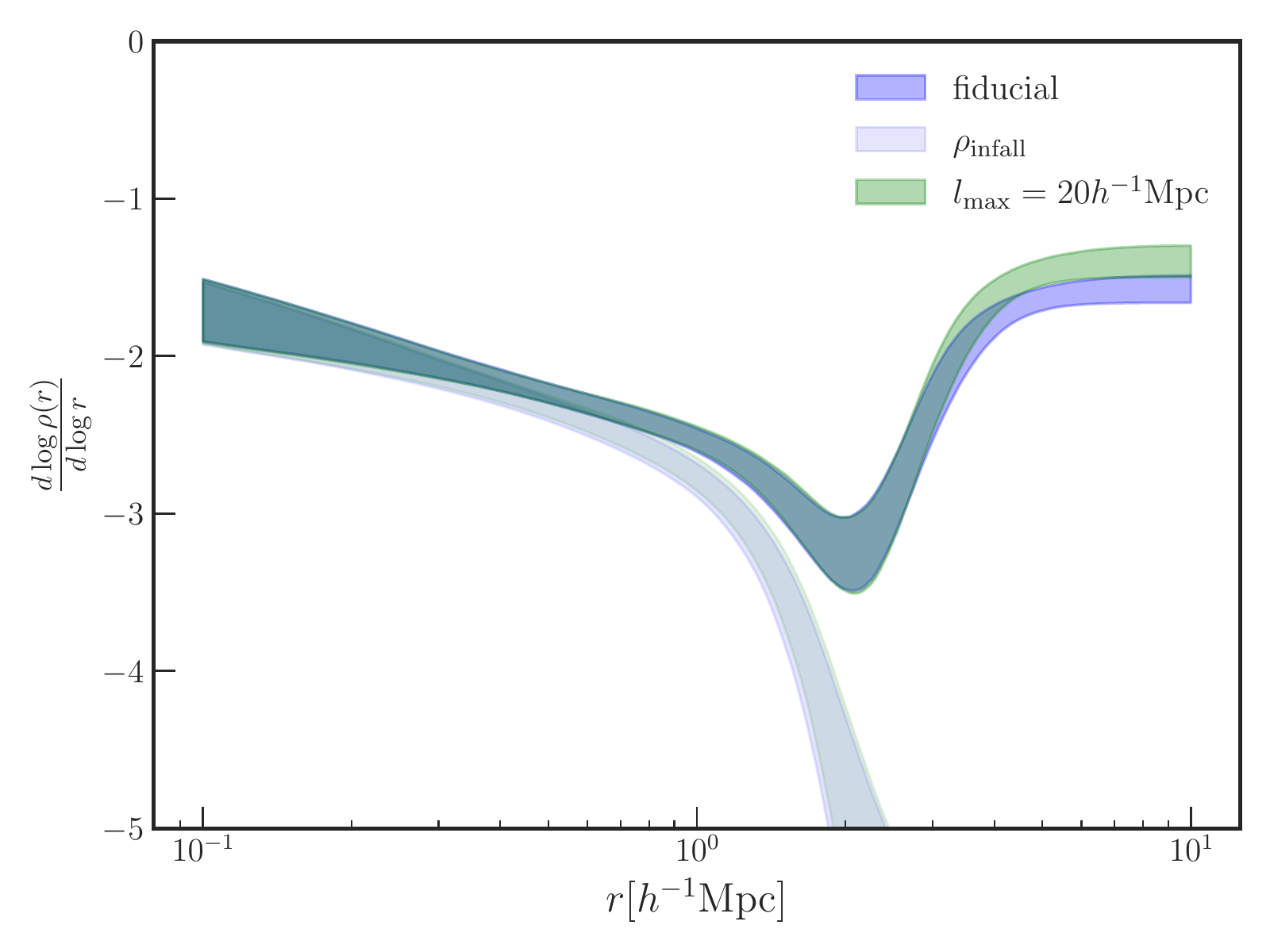}
\caption{The comparison of the fitted logarithmic slope profiles between the fiducial model (red) and the model with 20 $h^{-1} {\rm Mpc}$ projection length (blue).
}
\label{fig:slope_lmax}
\end{figure}

Finally, we test how much the projection length ($l_{\rm max}$) affects our fitting, by changing it to $l_{\rm max}=20 h^{-1} {\rm Mpc}$. 
The result is shown in the Fig.~\ref{fig:slope_lmax}.
While it does not alter our result significantly, we note that the outer slope gets altered by $\sim 1 \sigma$.
It suggests that one should use caution when interpreting the outer slope parameter $s_{\rm e}$.

\section{Covariance Matrices}
\label{app:cov}
\begin{figure}
\centering
\includegraphics[width=0.99\linewidth]{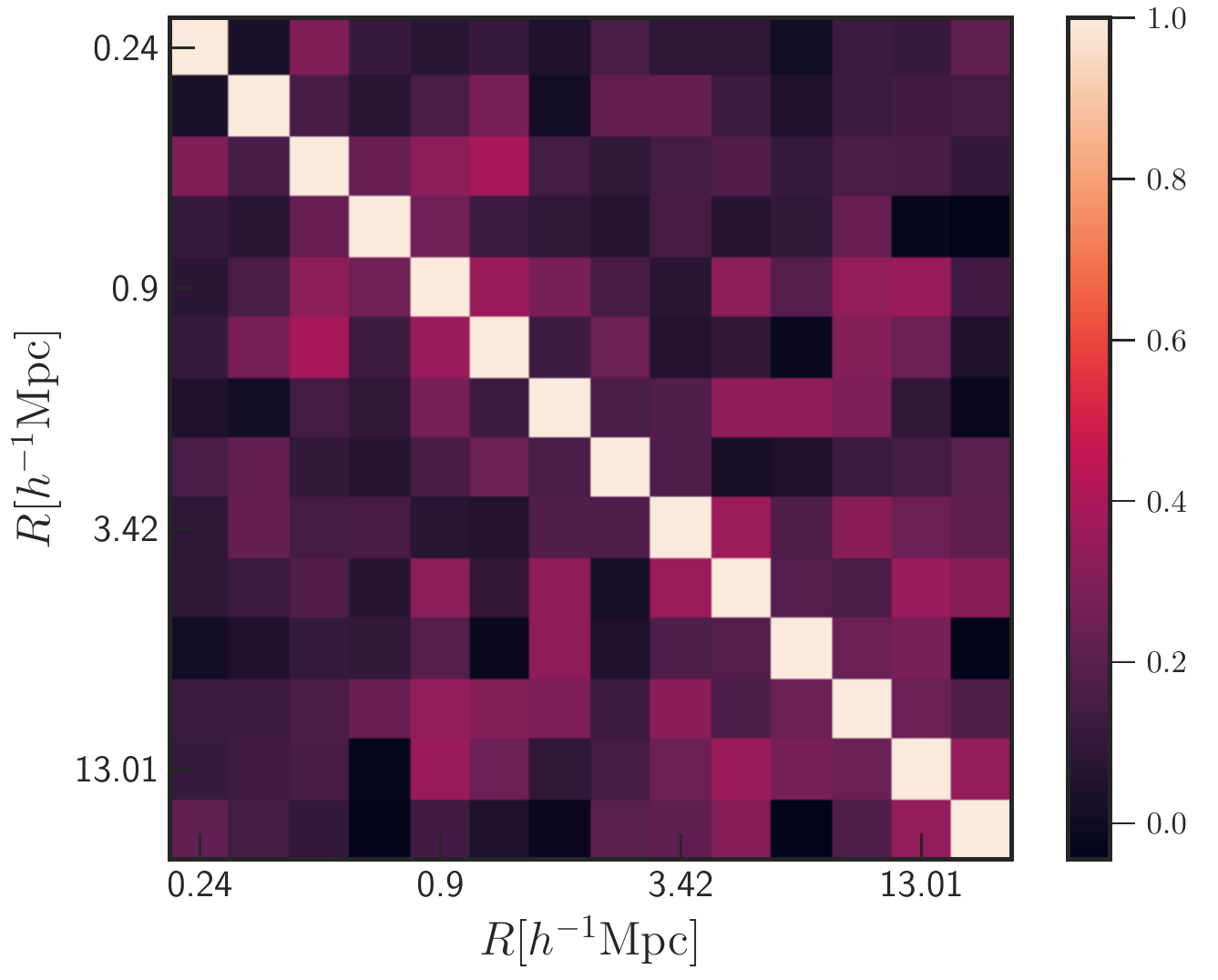}
\includegraphics[width=0.99\linewidth]{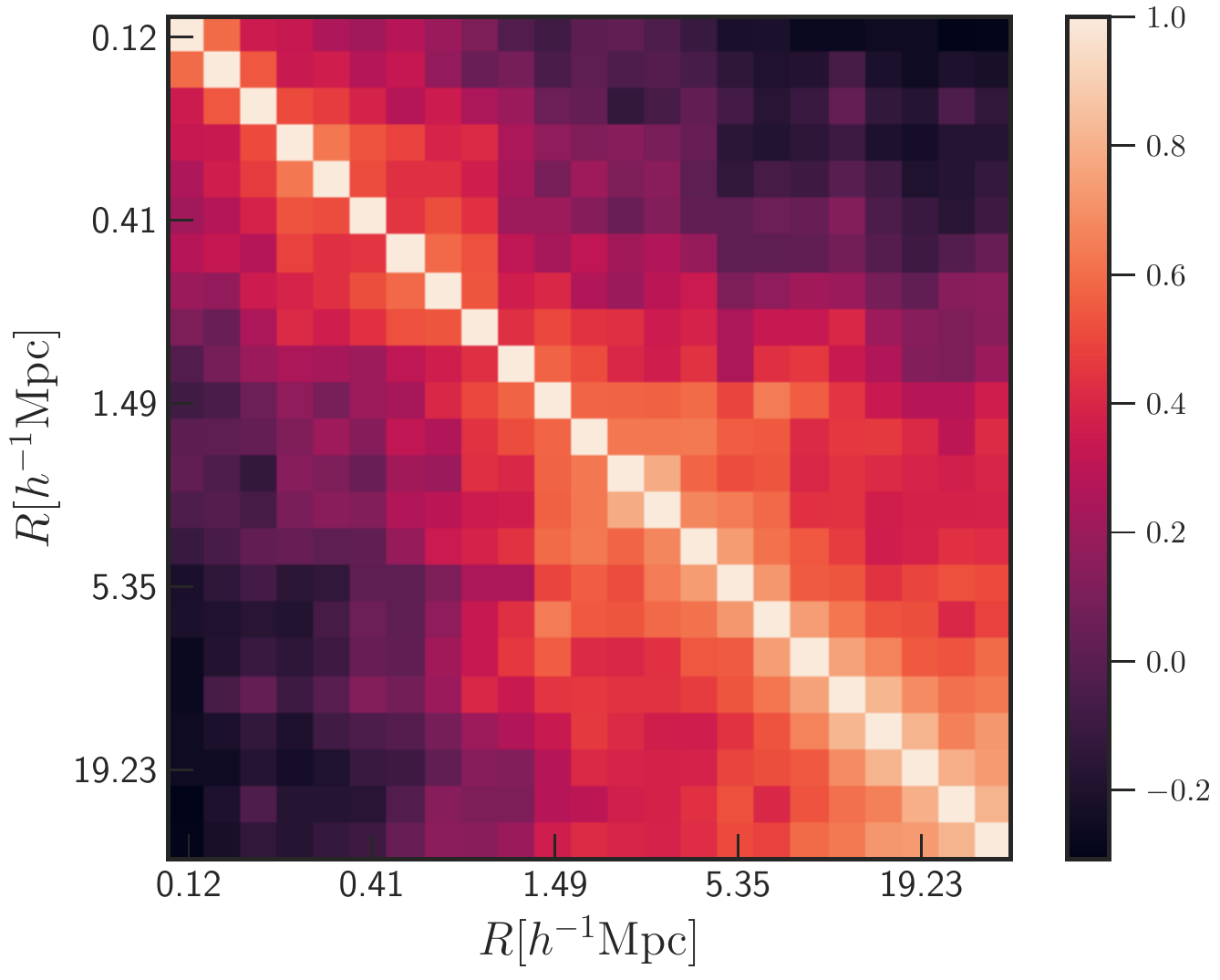}
\caption{
The normalized covariance matrices for $\Delta\Sigma(R)$ (top, see Sec.~\ref{sec:measure-dsigma}) and $\Sigma_{\rm g} (R)$ (bottom, see Sec.~\ref{sec:measure-sigmag}). 
Note that the covariance matrix for $\Delta\Sigma(R)$ includes the boost factor correction (see Sec.~\ref{sec:boost}).
}
\label{fig:cov_mat}
\end{figure}

In Fig.~\ref{fig:cov_mat}, we show the normalized covariance matrices (correlation matrices) for the WL profile ($\Delta\Sigma$, see Sec.~\ref{sec:measure-dsigma}) and the galaxy surface density profile ($\Sigma_{\rm g}$, see Sec.~\ref{sec:measure-sigmag}). 
Note that the covariance matrix for the $\Delta\Sigma$ includes the contribution from the boost factor correction (see Sec.~\ref{sec:boost}). It is close to diagonal as it is dominated by shape noise. 

\end{document}